\documentclass[american,aps,prxquantum,reprint,superscriptaddress,longbibliography,floatfix,nofootinbib,showkeys]{revtex4-2}
\usepackage[T1]{fontenc}
\usepackage[utf8]{inputenc}
\usepackage{color}
\usepackage{centernot}
\usepackage{babel}
\usepackage{physics}
\usepackage{mathtools}
\usepackage{bbding}
\usepackage{pifont}
\usepackage{textcomp}
\usepackage{wasysym}
\usepackage{amsthm}
\usepackage{nicefrac}
\usepackage{dsfont}
\usepackage{amsmath}
\usepackage{amssymb}
\usepackage{graphicx}
\usepackage{enumitem}
\usepackage{cancel}
\definecolor{darkblue}{rgb}{0.0, 0.0, 0.55}

\usepackage[colorlinks=true,
            allcolors=blue]{hyperref}

\hypersetup{
     colorlinks   = true,
     linkcolor    = blue,
     citecolor    = blue,
     urlcolor     = blue,
     }

\usepackage{appendix}
\renewcommand{\selectlanguage}[1]{}
\makeatletter
\@ifundefined{textcolor}{}
{%
	\definecolor{BLACK}{gray}{0}
	\definecolor{WHITE}{gray}{1}
	\definecolor{RED}{rgb}{1,0,0}
	\definecolor{GREEN}{rgb}{0,1,0}
	\definecolor{BLUE}{rgb}{0,0,1}
	\definecolor{CYAN}{cmyk}{1,0,0,0}
	\definecolor{MAGENTA}{cmyk}{0,1,0,0}
	\definecolor{YELLOW}{cmyk}{0,0,1,0}
}
\theoremstyle{plain}

\theoremstyle{plain}

\ifx\proof\undefined
\newenvironment{proof}[1][\protect\proofname]{\par
	\normalfont\topsep6\p@\@plus6\p@\relax
	\trivlist
	\itemindent\parindent
	\item[\hskip\labelsep
	\scshape
	#1]\ignorespaces
}{%
	\endtrivlist\@endpefalse
}
\theoremstyle{remark}
\newtheorem*{remark}{Remark}
\providecommand{\proofname}{Proof}
\fi
\theoremstyle{plain}

\providecommand{\lemmaname}{Lemma}
\providecommand{\definitionname}{Definition}
\providecommand{\propositionname}{Proposition}

\usepackage{txfonts}
\usepackage{colortbl}\definecolor{myurlcolor}{rgb}{0,0,0.7}

\def\ket#1{| #1 \rangle}
\def\bra#1{\langle  #1 |}

\def\proj#1{| #1 \rangle\!\langle #1 |}
\usepackage[normalem]{ulem}
\usepackage{tikz}
\usetikzlibrary{circuits.ee.IEC}
\usetikzlibrary{positioning, shapes, arrows.meta, fit, calc}
\usetikzlibrary{arrows.meta, decorations.pathmorphing, patterns}

\newtheorem{example}{Example}

\newtheorem{remark}{Remark}

\newtheorem{proposition}{Proposition}
\newtheorem{definition}{Definition}

\definecolor{orange}{RGB}{255,127,0}

\makeatletter
\renewcommand{\thesection}{\arabic{section}}
\renewcommand{\thesubsection}{\arabic{section}.\arabic{subsection}}

\renewcommand{\p@subsection}{}
\renewcommand{\p@subsubsection}{}
\makeatother

\renewcommand{\geq}{\geqslant}
\renewcommand{\le}{\leqslant}
\renewcommand{\ge}{\geqslant}

\begin{document}
\title{Genuine and Non-Genuine Quantum Non-Markovianity: A Unified Information-Theoretic Review}

\author{Rajeev Gangwar}
\email[Corresponding author: ]{rajeev@unist.ac.kr}
\affiliation{Department of Mathematics, Technion - Israel Institute of Technology, Haifa 3200000, Israel}
\affiliation{Department of Physics, Ulsan National Institute of Science and Technology (UNIST), Ulsan 44919, Republic of Korea}

\author{Ujjwal Sen}
\email{ujjwal@hri.res.in, ujjwalsen0601@gmail.com}
\affiliation{Harish-Chandra Research Institute, Chhatnag Road, Jhunsi, Prayagraj 211 019, India}
\affiliation{Homi Bhabha National Institute, Training School Complex, Anushakti Nagar, Mumbai 400 094, India}

\begin{abstract}
Understanding whether the features of open quantum dynamics are genuinely quantum remains a central challenge in the theory of open quantum systems. Even though the non-Markovian behavior of quantum dynamics has been widely investigated across different settings, there is still no consensus on which properties of a dynamics reflect genuine quantum features and which arise from classical or non-genuine quantum sources. Here, we review recent developments in characterizing quantum non-Markovianity based on information backflow and the nature of its origin. We also present a survey on how various approaches separate classical and quantum contributions, as well as how they define operational tasks that reveal genuine quantum non-Markovianity. We analyze several frameworks, including state-distinguishability-based, channel-based (``CP-divisibility''), and process-tensor methods. For each framework, we outline the underlying physical motivation and the criteria proposed to distinguish genuine quantum non-Markovianity from practical or apparent memory effects. We further compare different approaches and their strengths and limitations. The review aims to clarify the conceptual and operational aspects of quantum non-Markovian processes based on their nature and to provide a foundation for future research on quantum non-Markovianity and its role in advancing quantum information science
and technology.
\end{abstract}


\maketitle

\section{Introduction}
When analyzing the time evolution of physical systems, possibly tailor-made for information processing, a central question is whether the system’s future behavior depends solely on its present state or whether its past continues to influence its dynamics. This distinction is important because many real processes do not simply “forget” their history; instead, they carry remnants of earlier configurations that shape their subsequent evolution. Such retained information can arise from interactions with an environment, internal structure, or correlations built up over time. Understanding how these features affect the dynamics provides a natural motivation to understand the concept of non-Markovianity, which captures the role of memory.

\begin{figure}[h]
\centering
\resizebox{\columnwidth}{!}{%
\usetikzlibrary{arrows.meta,decorations.pathmorphing, patterns}

\begin{tikzpicture}[scale=0.50, transform shape,
  every node/.style={font=\small,inner sep=0,outer sep=0},
  cup/.style={draw=black,line width=0.6pt,fill=white},
  coffee/.style={fill=brown!70!black},
  loss/.style={-Stealth, thick, draw=orange!80!black},
  reduced/.style={-Stealth, thick, draw=orange!50!black, dashed, opacity=0.6},
  backflow/.style={Stealth-, thick, draw=blue!70!black, dotted, line width=1pt}
]

\begin{scope}[xshift=5.0cm, yshift=-1.8cm, scale=2, transform shape]

  \draw[line width=0.9pt] (-3.1,-0.9) rectangle (3.1,4.1);
  \begin{scope}
    \clip (-3.1,-0.9) rectangle (3.1,4.1);
    \fill[pattern=north east lines, pattern color=gray!40] (-3.1,-0.9) rectangle (3.1,4.1);
  \end{scope}

  \draw[line width=0.9pt, fill=white] (-2.6,-0.35) rectangle (2.6,3.65);
  \node[anchor=north west] at (-1.5,4.05) {Partially insulated wall};

  \begin{scope}[shift={(0,0.8)}]
    \fill[gray!10] (0,-0.6) ellipse (2cm and 0.42cm);
    \draw[line width=0.6pt] (0,-0.6) ellipse (2cm and 0.42cm);

    \path[fill=white] (-1.4,-0.6) -- (-1.1,1.0) arc(180:0:1.1 and 0.35) -- (1.4,-0.6);
    \draw[cup] (-1.4,-0.6) -- (-1.1,1.0) arc(180:0:1.1 and 0.35) -- (1.4,-0.6);

    \draw (0,1.0) ellipse (1.05cm and 0.32cm);
    \fill[coffee] (0,1.0) ellipse (0.95cm and 0.26cm);
  \end{scope}

  \draw[loss] (-0.7,1.8) .. controls (-0.8,2) and (-1,3) .. (-3.4,2.5);
  \draw[loss] (-0.4,1.6) .. controls (-0.7,2.2) and (-0.2,2.3) .. (-0.5,2.6);
  \draw[loss] (-0.1,1.6) .. controls (-0.2,2) and (0.1,2.12) .. (0.1,2.5);
  \draw[loss] (0.2,1.6) .. controls (0.2,2.2) and (0.4,2.3) .. (0.6,2.6);
  \draw[loss] (0.5,1.6) .. controls (0.6,2.2) and (0.6,2.4) .. (1.2,2.5);
  \draw[loss] (0.7,1.8) .. controls (0.8,2) and (1,3) .. (3.4,2.3);

  \node at (0,2.9) {Loss of heat};

  \draw[backflow] (-0.5,1.1) .. controls (-1.5,0) and (-1,2) .. (-2.6,1);
  \draw[backflow] (-0.5,1.5) .. controls (-1.5,0.5) and (-1,2.5) .. (-2.6,1.5);
  \draw[backflow] (0.5,1.5) .. controls (1.5,0.5) and (1,2.5) .. (2.6,1.5);
  \draw[backflow] (0.5,1.1) .. controls (1.5,0) and (1,2) .. (2.6,1);
  \node at (0,0.5) {Backflow of heat};

  \node[below=10pt] at (0,-0.7) {Non-Markovian dynamics};

\end{scope}

\end{tikzpicture}
}
\caption{\textbf{Non-Markovian dynamics:} A mental picture of a non-Markovian process can be a cup of hot tea kept inside a container having partially insulated walls (partially insulating to heat). The cup releases heat into the immediate environment (orange arrows), but because the walls are partially insulated, some of the heat flows back (blue lines) to the cup, creating a ``backflow'' and introducing memory into the process. Some heat still escapes through the insulation to the external, faraway environment (long orange arrows).
}
\label{fig:example_non-Markov}
\end{figure}
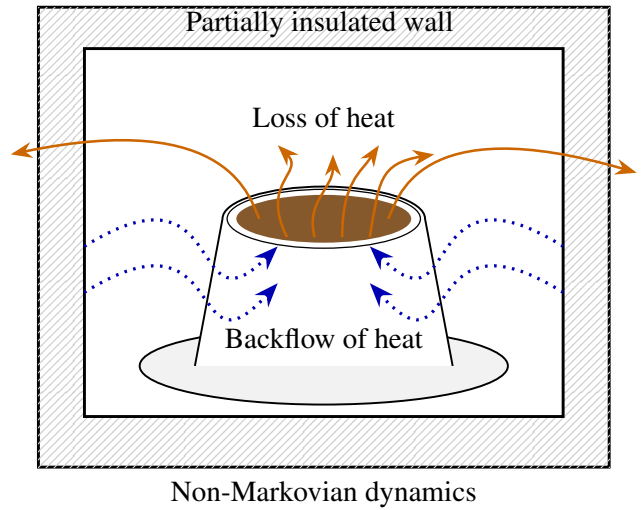

\begin{figure}[h]
\centering
\resizebox{\columnwidth}{!}{%
\usetikzlibrary{arrows.meta,decorations.pathmorphing,patterns}

\begin{tikzpicture}[scale=0.10, transform shape,
  every node/.style={font=\scriptsize, inner sep=0, outer sep=0},
  cup/.style={draw=black, line width=0.07pt, fill=white},
  coffee/.style={fill=brown!70!black},
  loss/.style={-{Stealth[length=0.3mm,
  width=0.2mm]},
draw=orange!80!black,
line width=0.085pt},
]

\begin{scope}[xshift=10cm]
\fill[gray!10] (0,-0.6) ellipse (2cm and 0.42cm);
\draw[line width=0.05pt] (0,-0.6) ellipse (2cm and 0.42cm);

\path[fill=white] (-1.4,-0.6) -- (-1.1,1.0)
    arc(180:0:1.1 and 0.35) -- (1.4,-0.6);
\draw[cup] (-1.4,-0.6) -- (-1.1,1.0)
    arc(180:0:1.1 and 0.35) -- (1.4,-0.6);

\draw[line width=0.08pt] (0,1.0) ellipse (1.05cm and 0.32cm);
\fill[coffee] (0,1.0) ellipse (0.95cm and 0.26cm);

\draw[loss] (-0.1,1) .. controls (-0.2,1.3) and (0.1,1.6) .. (0.1,2);
\draw[loss] (-0.4,1) .. controls (-0.7,1.4) and (-0.2,1.6) .. (-0.5,2.1);
\draw[loss] (0.2,1) .. controls (0.2,1.5) and (0.4,1.8) .. (0.6,2);
\draw[loss] (0.5,1) .. controls (0.6,1.1) and (0.6,1.65) .. (1,1.8);

\node at (0,2.3) {Loss of heat};
\node[below=10pt] at (0,-0.85) { Markov dynamics};
\end{scope}
\end{tikzpicture}
}
\caption{\textbf{Markov dynamics:}
We consider the same cup of hot tea as in Fig.~\ref{fig:example_non-Markov}, but this time without the partially-insulated bounding walls. The cup releases heat to the environment (orange arrows). The environment does not store or return this heat (there are no blue arrows this time), and so cooling occurs in a one-way, memoryless manner, with no heat backflow. This type of dynamics is referred to as Markovian.}
\label{fig:example_markov}
\end{figure}
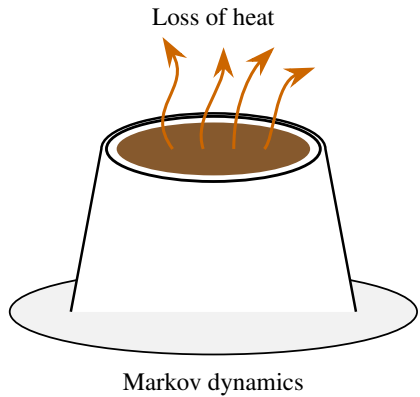

Non-Markovianity, in a general sense, refers to the presence of memory effects in the evolution of a system, independent of whether the system is classical or quantum. A process is non-Markovian when its future dynamics depend not only on its current state but also on its past history (for example, see Fig.~\ref{fig:example_non-Markov}). In such cases, the system retains information about earlier states, and this stored information influences its subsequent behavior. These memory effects give rise to temporal correlations that cannot be described by a simple memoryless Markovian process, where the evolution is fully determined by the present state alone (for example, see Fig.~\ref{fig:example_markov}). In nature, the future does not depend solely on the present; it typically also depends on the past. Non-Markovian dynamics, therefore, capture a broad class of processes in which the past plays an active role in shaping the future trajectory.
The relevance of non-Markovian processes
spans many natural and engineered systems. In statistical physics~\cite{Milz_2021}, non-Markovian models account for anomalous diffusion~\cite{DASILVA2015522,Vitali_2022,Barraza_2025,Chekroun_2025}, viscoelasticity relaxation~\cite{Volkov_1996,Ali_2023,azevedo2025}, and transport in disordered media~\cite{Armando_2021,alsabbagh2016,Iomin_2024}. In biological systems, memory effects appear in neural signaling~\cite{da_silva_2015,murty2024}, gene regulation~\cite{Yang_2021,Zhang_2020,Michael_2024}, and population dynamics~\cite{Ariel_2023}, where delayed feedback and temporal correlations are essential. In economics and finance, non-Markovian models capture long-range temporal correlations and volatility clustering in market behavior. These examples show the broad importance of non-Markov processes and motivate the development of general frameworks for open-system dynamics.

The study of non-Markovian dynamics is classified at the broad level into classical and quantum non-Markovianity, depending on the nature of the system and the type of information that flows back into it during evolution. Classical non-Markovianity arises when past states influence future behavior through classical correlations, which is a usual signature of memory in the dynamics.

In contrast, quantum non-Markovianity deals with memory effects that appear in the evolution of quantum states or quantum processes. Here, the system may become correlated with its environment in ways that cannot be captured by classical descriptions. Quantum coherence~\cite{Martin_2017}, entanglement~\cite{horodecki-2009-review,Das_2016}, and other quantum correlations~\cite{Vedral_2012,Bera_2018} that are stored in the environment may later return and influence the system’s future evolution, and understanding of non-Markovian dynamics is also very important for quantum error correction methods~\cite{wfyl-wtz3,29yv-12sq,Gangwar2023}. However, not all quantum non-Markovian behavior is genuinely quantum in general. Some memory effects observed in quantum systems may still be reproduced by effective classical models or arise from correlations that do not explicitly require quantum resources~\cite{Milz_2020,Breuer_2018-mixing-induced}. Such non-genuine contributions can be identified and separated using tools such as squashed quantum non-Markovianity~\cite{Gangwar2024,Buscemi_2025} and hysteretic squashed entanglement~\cite{das2026}.

\begin{figure}[h]
\centering
\resizebox{\columnwidth}{!}{%
\usetikzlibrary{arrows.meta, positioning}

\begin{tikzpicture}[
  node distance=15mm,
  box/.style={
    draw,
    rectangle,
    rounded corners=2pt,
    minimum width=4.2cm,
    minimum height=7mm,
    align=center,
    font=\large
  },
  arr/.style={
    -{Stealth[length=4mm]},
    line width=0.8pt
  }
]

\node[box] (NM) {Non-Markovianity};

\node[box, below=18mm of NM, xshift=-28mm] (C) {Classical\\Non-Markovianity};
\node[box, below=18mm of NM, xshift=28mm]  (Q) {Quantum\\Non-Markovianity};

\node[box, below=15mm of Q, xshift=-30mm] (G) {Non-genuine quantum\\non-Markovianity};
\node[box, below=15mm of Q, xshift=30mm]  (NG) {Genuine quantum\\non-Markovianity};

\draw[arr] (NM) -- (C);
\draw[arr] (NM) -- (Q);

\draw[arr] (Q) -- (G);
\draw[arr] (Q) -- (NG);

\end{tikzpicture}
}
\caption{Classification of non-Markovianity in quantum dynamics of physical systems.}
\label{fig:nonmarkov-hierarchy}
\end{figure}
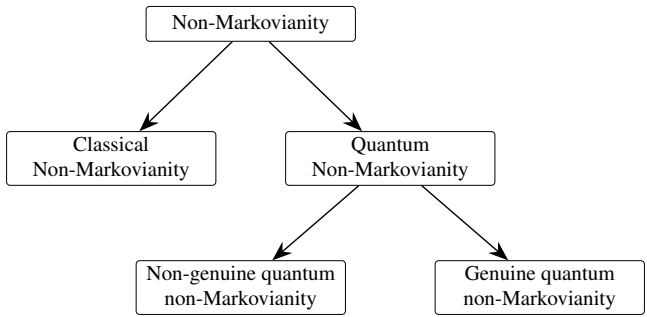

For this reason, quantum non-Markovianity is further divided into genuine and non-genuine parts. Genuine quantum non-Markovianity refers to memory effects that rely on intrinsically quantum features, such as entanglement-mediated feedback or quantum correlations that cannot be simulated by classical processes. These effects represent the truly quantum contribution to the temporal correlations in the dynamics. Non-genuine quantum non-Markovianity, on the other hand, includes memory effects that may occur in quantum systems, but which can be explained through classical correlations or effective classical noise models. This hierarchical structure clarifies the different origins of memory and allows one to distinguish classical memory effects from those arising solely from quantum effects, as illustrated schematically in Fig.~\ref{fig:nonmarkov-hierarchy}.

In this review, attention will be focused on how the genuine and non-genuine components of quantum non-Markovianity are addressed in recent research. The discussion will trace how current studies refine the boundary between these two classes, develop tools to characterize them, and explore their relevance in practical quantum information tasks. By outlining these developments, the review aims to give a clear picture of the progress made in understanding the structure of memory effects and the emerging directions that continue to shape this field.

\tableofcontents

\section{Classical non-Markovian dynamics}
An understanding of classical Markov and non-Markovian dynamics is essential before discussing quantum Markov and non-Markovian dynamics in open quantum systems and their applications in quantum information processing. Since our goal is to discuss quantum dynamics, we will briefly review classical Markov dynamics here.  Classical non-Markovianity has provided deep insights into the mechanisms of irreversibility, fluctuation-dissipation relations, and the emergence of non-equilibrium steady states, also discussed in the review article~\cite{Rivas_2014}. Understanding these memory-driven mechanisms is essential for modeling realistic complex systems that deviate from idealized Markovian assumptions and for developing predictive and control frameworks in stochastic dynamics~\cite{Milz_2021}.

\begin{definition}
\textbf{Classical Markovianity:} A stochastic process is known as a Markov process if the conditional probability that the random variable $X$ takes the value $x_n$ at time $t_n$, given its values at all earlier times $t_{n-1}>t_{n-2}>\cdots>t_0$, depends only on the most recent value $x_{n-1}$:
\begin{align}\label{classical_markov}
&P(x_n, t_n|x_{n-1}, t_{n-1}; \ldots ; x_0, t_0) = P(x_n, t_n|x_{n-1}, t_{n-1}),
\end{align}
for all $n$, all ordered times $t_n>t_{n-1}>\cdots>t_0$, and all values $x_n,\ldots,x_0$.
\end{definition}
In other words, the values taken by $X$ at times earlier than $t_{n-1}$ do not affect the complete probability distribution. It can be summarized by the statement that a Markov process does not retain any memory of the history of $ X$'s past values. Therefore, it is clear that in Markov dynamics, the past cannot influence future dynamics. This kind of stochastic process is named after the Russian mathematician A.~A.~Markov.  In addition, Markov dynamics also satisfy another property, such as if we take the joint probability distribution for any three consecutive times $t_3 > t_2 > t_1$ and use the definition of
conditional probability twice, then we obtain
\begin{align}
&P(x_3, t_3; x_2, t_2; x_1, t_1) \notag\\
&\quad= P(x_3, t_3|x_2, t_2; x_1, t_1)P(x_2, t_2; x_1, t_1) \notag \\
&\quad= P(x_3, t_3|x_2, t_2; x_1, t_1)P(x_2, t_2|x_1, t_1) P(x_1, t_1). \label{eq:joint3}
\end{align}
Since the Markov condition~\eqref{classical_markov} implies that
$P(x_3, t_3|x_2, t_2; x_1, t_1) = P(x_3, t_3|x_2, t_2)$,
by taking the sum over $x_2$ and
dividing by $P(x_1, t_1)$ on both sides, we obtain
\begin{align}
P(x_3, t_3|x_1, t_1) =
\sum_{x_2 \in X}
P(x_3, t_3|x_2, t_2)P(x_2, t_2|x_1, t_1), \label{eq:CK_eq}
\end{align}
which is known as the Chapman-Kolmogorov equation and is well known in classical dynamics. Note that the Chapman--Kolmogorov equation is a necessary but not a sufficient condition for Markovianity: there exist classical non-Markovian processes whose two-time transition probabilities satisfy Eq.~\eqref{eq:CK_eq}. Hence, memory is in general a multi-time property, a point that will reappear in the quantum setting.

\section{Quantum non-Markovianity} \label{Defs_Markov}
In this section, we outline the basic definitions of quantum non-Markovianity and the methods used to quantify it, followed by several approaches for witnessing it and the corresponding witness techniques. Before introducing these tools, it is always helpful to present a general description of quantum non-Markovianity. In open quantum dynamics, a system interacts with its surrounding environment, and this interaction can give rise to memory effects, also known as backflow of information. Based on these memory effects, in quantum non-Markovian dynamics, quantum information exchange between the system and its environment occurs, with information temporarily leaving the system and later returning. If this bidirectional exchange of information results in dynamics that deviate from the memoryless (Markovian) picture, it provides a signature of non-Markovianity in quantum dynamics. Here, we will discuss and organize various measures and witnesses developed to characterize it. Whether this exchange of information has a genuinely quantum origin, such as coherence, entanglement, or more general quantum correlations, or can be explained classically, is not decided by these measures; this question is discussed in Sec.~\ref{sec:QC_QNM}.

There are two widely discussed approaches to describing quantum non-Markovianity: CP-divisibility and quantum state distinguishability, based on the Rivas–Huelga–Plenio (RHP)~\cite{Rivas_2010} and Breuer–Laine–Piilo (BLP)~\cite{Breuer2009} frameworks, respectively. We will begin with the RHP and BLP approach. After introducing these two frameworks, we discuss other definitions that have appeared in the literature, providing a broader view of the various ways in which non-Markovian behavior has been characterized.

\subsection{CP-divisibility: RHP quantum non-Markovianity}\label{subsec:RHP_QNM}
CP-divisibility provides one of the most widely used structural characterizations of quantum non-Markovianity. In this framework, deviations from complete positivity of intermediate dynamical maps signal the presence of memory effects in open-system dynamics. According to the CP-divisibility approach~\cite{Rivas_2010}, Markovian dynamics is defined as follows.
\begin{definition}[RHP non-Markovianity]
A family of CPTP dynamical maps $\{\Lambda_t\}_{t\ge 0}$ is said to be RHP-Markovian if it is CP-divisible.
In other words, for all times $t \ge s \ge 0$, the evolution can be written as
\begin{align}
    \Lambda_t = V_{t,s}\,\Lambda_s,  \label{eq:CP-div}
\end{align}
where $V_{t,s}$ is a completely positive and trace-preserving (CPTP) intermediate map. If $\Lambda_s$ is invertible, the intermediate propagator
\begin{align}
    V_{t,s} = \Lambda_t \circ \Lambda_s^{-1}
\end{align}
exists and is unique, and CP-divisibility is equivalent to it being completely positive for all $t \ge s$.
\end{definition}
If such a decomposition does not exist, the process is RHP non-Markovian. In this approach, they have also introduced an information-theoretic approach to detect and quantify non-Markovianity by examining the evolution of quantum correlations between a system and an isolated ancilla. Their method relies on the fact that entanglement cannot increase under any CP-divisible evolution; therefore, any temporary rise or revival of correlations between the system and ancilla indicates the presence of memory effects. Within this framework, they propose two measures: one that is necessary and sufficient when the full dynamical map's information is available, and another that provides an adequate indicator when only partial knowledge of the dynamics is accessible.

In the setting where no prior information about the dynamics is assumed, the procedure begins with the preparation of a maximally entangled state between the system and the ancilla,
\begin{align}
    |\phi^+\rangle = \frac{1}{\sqrt{d}} \sum_{n=0}^{d-1} |n\rangle |n\rangle, \qquad
    \rho_{SA} = |\phi^+\rangle\langle\phi^+|.
\end{align}
Under a CP-divisible evolution, the reduced system undergoes a local CPTP map between any two times, which ensures that any valid entanglement measure $E$ (i.e., any entanglement monotone under local operations) decreases monotonically in time. Deviations from this monotonic behavior signal the presence of RHP non-Markovianity in the dynamics. To quantify such deviations, the change in entanglement between an initial time $t_0$ and a final time $t_{\max}$ is defined as
\begin{align}
    \Delta E = E[\rho_{SA}]_{t_0} - E[\rho_{SA}]_{t_{\max}}.
\end{align}
The corresponding non-Markovianity measure is given by
\begin{align}
    I &=
    \int_{t_0}^{t_{\max}}
    \left| \frac{d}{dt} E[\rho_{SA}(t)] \right| dt
    - \Delta E \nonumber\\
    &=2\int_{\frac{d}{dt}E>0}\frac{d}{dt} E[\rho_{SA}(t)]\, dt\;\ge\;0 .
\end{align}
The second equality follows by splitting the time interval into the parts where
the entanglement increases and where it decreases. Writing
$P=\int_{\frac{d}{dt}E>0}\frac{d}{dt}E\,dt\ge 0$ and
$N=\int_{\frac{d}{dt}E<0}\big|\frac{d}{dt}E\big|\,dt\ge 0$, one has
$\int_{t_0}^{t_{\max}}\big|\frac{d}{dt}E\big|\,dt=P+N$, whereas
$\Delta E=-\int_{t_0}^{t_{\max}}\frac{d}{dt}E\,dt=N-P$. Hence $I=2P$.
For CP-divisible evolutions, the derivative $\tfrac{d}{dt} E[\rho_{SA}(t)]$ is non-positive, which implies $\int|\tfrac{d}{dt}E|\,dt=\Delta E$ and hence $I=0$. Any strictly positive value $I>0$, therefore, indicates an increase in entanglement during the dynamics, signaling RHP non-Markovian behavior.

However, later analyses have shown that the relation between CP-divisibility and memory is subtle. First, CP-divisibility excludes information backflow for both invertible and non-invertible maps (by the contraction argument of Sec.~\ref{sec:Comparison}), but for non-invertible maps, the converse statement requires additional conditions when information is probed by pairwise trace distances~\cite{Erling_2018}, but not when arbitrary ensembles with an ancilla are used~\cite{Buscemi2016a}. Second, CP-divisibility of the two-time maps does not imply Markovianity of the full multi-time process~\cite{Milz2019} (Sec.~\ref{subsec:process_QNM}). This distinction becomes important when assessing non-Markovianity in scenarios where the dynamical maps may lose invertibility during the evolution. This observation underscores that divisibility-based notions of Markovianity must be interpreted with caution, particularly in physically relevant situations where the loss of invertibility obscures~\cite{Erling_2018} the direct connection between the structural properties of the map and the operational signatures of memory.

A recent work~\cite{Nakagawa2026} recasts the CP-divisibility test in the language of quantum characteristic functions, with the dynamical map itself, rather than a state, as the object being characterized. The Choi state of an intermediate propagator $V_{t,s}$ is expanded in a unitary operator basis, which defines a characteristic function and an associated Gram matrix. By a Bochner-type argument, this matrix is positive semidefinite if and only if the Choi operator is positive, so complete positivity of $V_{t,s}$, and hence RHP-Markovianity, is equivalent to positivity of this Gram matrix at all pairs of times $t\ge s$. In examples with time-dependent decay rates that temporarily become negative (amplitude damping and pure dephasing), the Gram matrix develops a negative eigenvalue in the windows where CP-divisibility fails (i.e., where the Choi operator has a negative eigenvalue); for amplitude damping these windows coincide with revivals of the trace distance, and for dephasing with revivals of coherence.

Conceptually, this approach is closely related to the RHP criterion: the Gram matrix carries the same information as the spectrum of the Choi operator, and it therefore relies on the same assumptions, including an invertible $\Lambda_s$ and two-time maps. As formulated, it detects CP-indivisibility; certifying information backflow, or distinguishing classical from quantum memory, would require additional structure. The agreement with trace-distance revivals in these qubit examples reflects the coincidence of CP- and $P$-divisibility there (Sec.~\ref{sec:Comparison}). An appealing feature of this formulation is that it places divisibility-based Markovianity in the positive-type, characteristic-function form familiar from quantum statistics, which may suggest extensions to multi-time processes.

\subsection{Distinguishability: BLP quantum non-Markovianity}\label{subsec:BLP_QNM}
The Breuer--Laine--Piilo (BLP) approach provides an operational characterization of non-Markovianity based on the distinguishability of quantum states. Unlike divisibility-based definitions, this framework focuses on how the ability to discriminate between states evolves under open-system dynamics. In this picture, non-Markovianity is identified with a temporary recovery of distinguishability, interpreted as information flowing back from the environment to the system. According to the BLP framework~\cite{Breuer2009}, non-Markovianity is defined as follows.
\begin{definition}[BLP non-Markovianity]
A quantum process described by a family of CPTP maps $\{\Lambda_t\}_{t\ge 0}$ is called BLP Markovian if the trace distance between any two evolving states
$\rho_1(t)=\Lambda_t(\rho_1(0))$ and $\rho_2(t)=\Lambda_t(\rho_2(0))$ decreases monotonically in time, i.e.,
\begin{align}
\frac{d}{dt} D\Big[\rho_1(t),\rho_2(t)\Big] \le 0 \quad \quad \forall\, t ,
\end{align}
\end{definition}
where the trace distance is defined as
\begin{align}
D[\rho_1,\rho_2]=\tfrac{1}{2}\|\rho_1-\rho_2\|_1 .
\end{align}
This monotonic decrease represents the contractive property of the trace distance under CPTP maps, reflecting a loss of distinguishability due to the system's interaction with its environment. In this setting, any increase in the trace distance would violate the divisibility property (specifically, CP-divisibility or $P$-divisibility, which will be discussed later in Sec.~\ref{sec:Comparison}) associated with an RHP Markovian evolution. If there exist time intervals during which
\begin{align}
\frac{d}{dt} D\Big[\rho_1(t),\rho_2(t)\Big] > 0
\end{align}
for at least one pair of initial states, the process is classified as a BLP non-Markovian process. Such a growth in distinguishability indicates a breakdown of monotonicity, signaling a temporary flow of information from the environment to the system. According to the BLP framework, this behavior is known as \emph{information backflow} from the environment to the system.

The BLP measure quantifies the total increase of distinguishability over all time intervals where the trace distance grows and is defined as
\begin{align}
\mathcal{N}_1
= \max_{\rho_1,\rho_2}
\int_{\sigma(t)>0} \sigma(t) dt ,
\end{align}
where
\begin{align}
\sigma(t)=\frac{d}{dt} D\Big[\rho_1(t),\rho_2(t)\Big].
\end{align}
A BLP Markovian evolution yields $\mathcal{N}_1=0$ since no positive contributions occur, while any revival of the trace distance leads to $\mathcal{N}_1>0$.

\begin{remark}
    Both the RHP and BLP definitions of quantum non-Markovianity capture the dynamical behavior in open-system dynamics, but they are not equivalent. Since positive trace-preserving maps contract the trace distance, any trace-distance revival implies that no positive (hence no completely positive) intermediate map exists: BLP non-Markovianity always implies RHP non-Markovianity. The converse fails. A clear illustration of this mismatch is provided in Appendix~\ref{exmpl_RHP-BLP} as an example. This example demonstrates that CP-indivisibility (RHP non-Markovianity) does not, in general, imply information backflow (BLP non-Markovianity): the dynamics violate CP-divisibility at all times, yet no revival of distinguishability occurs at any moment. Consequently, the model provides a clear and pedagogically valuable counterexample showing that the RHP and BLP notions of non-Markovianity capture fundamentally different physical properties of open-system dynamics.
\end{remark}

The idea of information flowing back to the system later encouraged the study of non-Markovianity not only through increases in distinguishability but also through the revival of correlations between the system and the environment using a reference system. These correlations have been described using mutual information~\cite{Luo2012}, conditional mutual information~\cite{Huang2021}, entanglement~\cite{Srana_2020,das2018fundamental}, interferometric power~\cite{Dhar2015,Souza2015}, and other related measures such as the temporal steering-based quantifier of Refs.~\cite{Chen2016,das2018fundamental}, which is discussed in Sec.~\ref{sec:temporal_steering}.

\subsection{Correlation: LFS quantum non-Markovianity} \label{sec:correlation_based}
The Luo-Fu-Song (LFS) approach is correlation-based to define the non-Markovianity~\cite{Luo2012}. The open-system dynamics of a quantum system $S$ is examined by introducing an ancilla $A$ that does not interact with the environment. The system evolves under a completely positive and trace-preserving dynamical map $\Lambda_t$, and the joint state at time $t$ is $\rho_{AS}(t) = (\mathds{1}_A \otimes \Lambda_t)\rho_{AS}(0)$. The central idea is that the total correlation shared between $A$ and $S$ reveals the memory properties of the dynamics. These correlations are quantified by the quantum mutual information, defined as
\begin{align}
I(A:S)_t = S(\rho_A(t)) + S(\rho_S(t)) - S(\rho_{AS}(t)),
\end{align}
where $S(\cdot)$ is the von Neumann entropy~\cite{nielsen_chuang_2010}, and $\rho_A(t)$ and $\rho_S(t)$ are the reduced states of system $A$ and $S$ obtained from $\rho_{AS}(t)$ after partial tracing. For a memoryless (Markovian) evolution, information flows only from the system to the environment; therefore, any correlations initially present between $A$ and $S$ can only decay with time. This leads to the monotonicity condition
\begin{align}
    \frac{d}{dt} I(A:S)_t \le 0 \quad \forall t . \label{DPI_vio}
\end{align}
A violation of the above inequality signals the revival of correlations. Such a violation amounts to a violation of the data processing inequality (DPI)~\cite{Buscemi2014} for the intermediate map: mutual information cannot increase under a local CPTP map on $S$, so an increase implies that the intermediate map is not CPTP, and that some information previously lost to the environment becomes accessible to $S$ again. Thus, the dynamics is classified as non-Markovian whenever there exist time intervals during which $\frac{d}{dt} I(A: S)_t > 0$.
\begin{definition}
A dynamical map $\Lambda_t$ is LFS non-Markovian whenever the mutual information between system $S$ and ancilla $A$ increases during the evolution, i.e.,
\begin{align}
    \frac{d}{dt} I(A\!:\!S)_t > 0.
\end{align}
\end{definition}
To quantify these revivals of correlations, in~\cite{Luo2012}, the authors introduce a measure based on the total amount of correlation revival. Let $\gamma(t) = \frac{d}{dt} I(A: S)_t$ denote the instantaneous rate of change of mutual information. The non-Markovianity measure of the dynamical map is defined as
\begin{align}
\mathcal{N}_2 = \max_{\rho_{AS}(0)} \int_{\gamma(t) > 0} dt\, \gamma(t),
\end{align}
where the maximization is performed over all initial $\rho_{AS}$ states. The integral accumulates only those time intervals in which the mutual information increases, so the measure quantifies precisely how much correlation is regained during the evolution. In this way, the correlation-based definition identifies non-Markovianity as the departure from the monotonic decay of total correlations, thereby providing an operationally meaningful measure grounded in information-theoretic principles. In addition to this measure, several studies have further explored information revival within this correlation-based perspective by considering environments with specific physical structures or by employing different types of correlations. These include approaches based on quantum discord~\cite{Alipour2012}, the concept of assisted knowledge~\cite{Fanchini2014}, decoherence-based methods~\cite{Haseli2014}, and more general classes of correlation measures~\cite{Santis2020}. Such works strengthen the interpretation that revivals of correlations, beyond total mutual information, serve as clear indicators of memory effects in open quantum dynamics.

\subsection{Conditional correlations-based non-Markovianity}

To gain a deeper understanding of this non-Markovian behavior driven by correlations, it is useful to analyze how correlations are redistributed among the ancilla, system, and environment during evolution. In this context, the quantum conditional mutual information naturally emerges as a refined quantity that captures changes in correlations beyond those visible in the mutual information approach. This leads to a characterization of quantum non-Markovianity based on conditional correlations. In this framework, a system $S$ interacts with an environment $E$, while an ancilla $A$ evolves trivially (via the identity channel). The joint state $\rho_{ASE}(t)$ evolves unitarily ($U_{SE}$), and the mutual information $I(S;A)$ is guaranteed to decrease monotonically if the reduced dynamics is CP-divisible~\cite{Luo2012}. However, in~\cite{Huang2021}, the authors demonstrate that this measure can be reformulated in terms of the quantum conditional mutual information (QCMI), which provides a clearer structural and information-theoretic characterization of the revival of correlations.

The key observation is that the total correlation between $A$ and the combined system–environment state is conserved: $I(A;SE)$ remains constant in time because the global evolution $U_{SE}$ is unitary while $A$ evolves through the identity channel. Using the chain rule of QCMI,
\begin{align}
    I(A;SE)=I(A;E|S)+I(S;A),
\end{align}
one finds that any increase in $I(S;A)$ must be balanced by a corresponding decrease in $I(A;E|S)$. Since $I(A;E|S)\ge 0$ by strong subadditivity, a negative derivative $dI(A;E|S)/dt<0$ signals that information that had leaked from the system into the environment becomes accessible to $S$ again.
\begin{definition}
A quantum process is non-Markovian whenever the quantum conditional mutual information
$\,I(A;E|S)_t\,$ decreases during the evolution, i.e.,
\begin{align}
    \frac{d}{dt} I(A;E|S)_t < 0.
\end{align}
\end{definition}
The corresponding quantifier of non-Markovianity, based on the total decrease of the QCMI, is given by
\begin{align}
\mathcal{N}(\Lambda)
=\sup_{\rho_{SA}} \int_{\frac{d}{dt}I(A;E|S)<0}
\left| \frac{d}{dt}I(A;E|S) \right| dt,
\end{align}
where the supremum is taken over all initial system–ancilla states. The integrand quantifies the instantaneous rate at which the conditional mutual information decreases, representing the amount of ``leaked information'' that flows back from the environment into the system. Because $I(A;E|S)$ measures how much of the environment's correlation with $A$ cannot be explained by the system alone, its reduction indicates that the system regains correlations or distinguishability resources that had previously propagated into the environment.

This CMI-based measure is fully equivalent to the LFS measure~\cite{Luo2012}. It therefore captures the same operational notion of non-Markovianity. Still, it does so by directly quantifying the decay of QCMI, which has established connections to recovery maps~\cite{petz2007quantum}, approximate Markov chains, and bounds on reconstructability~\cite{Hayden_2004}. The measure thus provides a structural and information-theoretic description of quantum non-Markovianity in terms of conditional correlations among ancilla, system, and environment.

\subsection{General contractivity-based non-Markovianity}
\label{subsec:Generalized contractive}
A broader perspective on non-Markovianity emerges when one studies how general contractive quantities behave under open-system dynamics, because contractive functions (See Appendix~\ref{sec:pre}) cannot increase under any CPTP map, for example, any contractive functions such as distinguishability measures, trace distance~\cite{Breuer2009}, Bures distance~\cite{Walter_1979}, but also correlation measures~\cite{Luo2012,Huang2021} which obey monotonicity under local operations. In CP-divisible evolutions, all such quantities must decrease monotonically; therefore, any temporary increase serves as an indicator of the backflow of information in the dynamics. In~\cite{Srana_2020}, this idea was examined and extended into a comprehensive, universal framework for witnessing non-Markovianity. The work addresses a fundamental question: Is the violation of monotonicity of any contractive function enough to detect every non-Markovian process? They show that contractive functions defined only on the system cannot capture all memory effects. For instance, for the eternally non-Markovian qubit dynamics of Appendix~\ref{exmpl_RHP-BLP}, no contractive function on the system alone detects non-Markovianity~\cite{Srana_2020}, and further subtleties arise for non-invertible evolutions~\cite{Andrzej2011,Bylicka_2017,Erling_2018}.

To resolve this limitation, in~\cite{Srana_2020}, the authors consider contractive functions and define an extended system ($A$) with an initial state that may be correlated with the system ($S$), while only the system interacts with the environment via a CPTP map. Following Refs.~\cite{Bae_2016, Buscemi2016a, Srana_2020}, non-Markovianity can be characterized as follows,

\begin{definition}\label{Def:4}
The dynamics $\{\Lambda_t\}$ is Markovian if, for every contractive function $f$, every ancilla $A$, every pair of joint states $\rho_{SA},\sigma_{SA}$, and all times $t\ge s\ge 0$,
\begin{align}
    &f\!\left[(\Lambda_t \otimes \mathds{1})(\rho_{SA}),\,(\Lambda_t \otimes \mathds{1})(\sigma_{SA})\right]\nonumber \\
    &\hspace{0.5cm}\le f\!\left[(\Lambda_s \otimes \mathds{1})(\rho_{SA}),\,(\Lambda_s \otimes \mathds{1})(\sigma_{SA})\right].\label{eq:contractive}
\end{align}
\end{definition}
The forward implication holds for every contractive $f$: CP-divisibility implies that all such quantities decay. The converse needs only one well-chosen quantity, and in the ensemble form of Ref.~\cite{Buscemi2016a} it holds without any invertibility assumption.

Under CP-divisible dynamics, this contractivity ensures a monotonic decrease of $f$ for all possible initial system and ancilla states, so if the dynamics satisfy Eq.~\eqref{eq:contractive}, then it will be Markovian dynamics, and the converse also holds for any non-Markovian dynamics; there always exists a pair of initial system and ancilla states and a suitable contractive function for which the monotonicity is violated. Therefore, an increase in a contractive function on the extended space is a universal signature of non-Markovianity: with general ensembles (guessing probability) it holds for arbitrary dynamics~\cite{Buscemi2016a}, while with pairwise quantities such as the trace distance it holds for invertible and image non-increasing dynamics and for all qubit dynamics~\cite{Bae_2016,Srana_2020}. A significant example analyzed in~\cite{Srana_2020} is entanglement negativity. Although bipartite system–ancilla negativity fails to detect some non-Markovian evolutions, the authors show that adding a further qubit $C$ changes the situation: for any invertible non-Markovian evolution, and for all (also non-invertible) qubit evolutions, there exists a tripartite state $\rho_{ABC}$ whose negativity across the $AB|C$ cut increases in time. Whether this extends to non-invertible dynamics beyond qubits is left open. This establishes negativity as a practical and universal indicator of non-Markovianity, capable of detecting memory even when single-system indicators, such as the trace distance, fail. These observations lead to a unified interpretation of quantum non-Markovianity in open quantum dynamics. This viewpoint generalizes the intuition behind both the RHP and BLP approaches: distinguishability-based signatures and correlation-based signatures become two aspects of the same principle when framed in terms of contractive quantities. It also highlights the importance of ancilla-assisted constructions, which reveal subtle memory effects that remain invisible in the system alone.

\subsection{Process tensors-based non-Markovianity}
\label{subsec:process_QNM}
Two-time dynamics capture important signatures of memory effects. However, two-time pictures are insufficient to describe the full temporal structure of open quantum dynamics (see example in Appendix~\ref{exam:exmp_pro_ten}). Memory is inherently a multi-time phenomenon. This motivates a complete characterization of temporal correlations in open quantum dynamics, which requires a multi-time description. In the process-tensor framework introduced by Pollock \emph{et al.}~\cite{Pollock2018PRA,Pollock2018} (see also Ref.~\cite{Milz2019}), the evolution of a system over times $t_0 < t_1 < \cdots < t_n$ is represented by the process tensor $T_{n:0}$, which is the multi-time Choi operator encoding all possible correlations between the system and its environment. The process tensor framework enables efficient computation via a time-translation-invariant matrix product operator representation~\cite{garbellini2026}. A quantum process is Markovian if its multi-time structure contains no temporal correlations beyond those generated by successive one-step CPTP maps.
This property is expressed directly at the level of the process tensor.
\begin{definition}[Markovian process tensor]
    A process is Markovian when the $n$-time process tensor $T_{n:0}$ factorizes into a tensor product of one-step Choi matrices describing the individual dynamical maps between consecutive times:
\begin{align}
T_{n:0}^{\mathrm{Markov}}
    = C_{t_n:t_{n-1}}
      \otimes C_{t_{n-1}:t_{n-2}}
      \otimes \cdots
      \otimes C_{t_1:t_0}.
\label{MarkovFactorization}
\end{align}
\end{definition}
Here each $C_{t_k:t_{k-1}}$ is the Choi matrix of a CPTP map acting during the interval $[t_{k-1},t_{k}]$. Eq.~\eqref{MarkovFactorization} is the defining property of a Markovian quantum process within the process-tensor formalism. It ensures that the process has no multi-time correlations: the dynamics at each step depend only on the system state at the immediately preceding time and not on any earlier history. All higher-order temporal correlations vanish, and the collection of one-step maps completely specifies the whole dynamics.

A widely used two-time criterion for Markovianity is CP-divisibility. A family of maps satisfying $\Lambda_{t_2:t_0} = \Lambda_{t_2:t_1}\, \Lambda_{t_1:t_0}$ with CPTP propagators is CP-divisible, but this is not sufficient for multi-time Markovianity~\cite{Milz2019} (see example in Appendix~\ref{exam:exmp_pro_ten}). Even if all propagators are CPTP, the full process tensor may still contain higher-order temporal correlations such that
\begin{align}
T_{n:0}
\neq
C_{t_n:t_{n-1}} \otimes \cdots \otimes C_{t_1:t_0}.\label{eq:process_NM}
\end{align}
Thus, CP-divisibility of the two-time maps does not imply that the process is Markovian in the multi-time sense, whereas Markovianity is fundamentally a multi-time concept.

\subsection{Approximate quantum non-Markovianity}
In this subsection, we discuss an approximate notion of Markovianity and the corresponding measure of non-Markovianity. Consider a system $S$ interacting with its immediate environment $E$, where the composite system $SE$ is further embedded in a larger environment $E_1$. The global system $SEE_1$ evolves unitarily, and for any initial product state $\rho_S^0 \otimes \rho_E^0 \otimes \rho_{E_1}^0$, the reduced joint dynamics of $SE$ is described by using the Stinespring dilation theorem~\cite{nielsen_chuang_2010},
\begin{align}
\rho_{SE}(t) = \Lambda_{SE}(\rho_S^0 \otimes \rho_E^0).
\end{align}
The construction of Markovianity in this framework uses the quantum mutual information, as in the LFS measure~\cite{Luo2012}, but evaluated between $S$ and $E$ rather than between $S$ and an ancilla, so that it quantifies the total correlation generated between the system and the environment during the evolution. A reduced map $\Lambda_S$ is called Markovian-like when no correlation is ever created between $S$ and $E$, namely when $I(S;E) = 0$ for all $t$ and for all initial product states. This notion is relaxed in Ref.~\cite{Das_2021}, where the authors define an $\epsilon$-Markovian map: for a fixed non-negative $\epsilon$, the map $\Lambda_S$ is said to be $\epsilon$-Markovian if
\begin{align}
I(S;E) \le \epsilon,
\end{align}
holds for every time $t$ and every initial product state. The set of all such maps is denoted by $\mathcal{S}_\epsilon$ and is referred to as the set of approximate Markov maps (the $\epsilon$-Markovian maps).
\begin{definition}
A map is $\epsilon$-Markovian when the generated system–environment correlation satisfies
\begin{align}
    I(S;E) \le \epsilon \quad \forall\, t.
\end{align}
\end{definition}
Thus, correlations may develop only within the controlled bound $\epsilon$.
This provides an approximate notion of Markovianity, which reduces to the exact Markovian-like case in the limit $\epsilon = 0$.

To quantify the deviation of an arbitrary dynamical map from this nearly Markovian behavior, Ref.~\cite{Das_2021} introduces the notion of $\epsilon$-non-Markovianity. Given a map $\Lambda_S$, its $\epsilon$-non-Markovianity is defined as the minimal distance between $\Lambda_S$ and the maps from the set $\mathcal{S}_\epsilon$,
\begin{align}
\mathcal{N}_\epsilon(\Lambda_S)=\min_{\widetilde{\Lambda}\in\mathcal{S}_\epsilon}\mathbf{D}(\Lambda_S,\widetilde{\Lambda}),
\end{align}
with
\begin{align}
\mathbf{D}(\Lambda, \widetilde{\Lambda}) =
\max_{\rho_S^0}
\mathbf{D}\left(\Lambda(\rho_S^0),\, \widetilde{\Lambda}(\rho_S^0)\right),
\end{align}
where, on the right-hand side, $\mathbf{D}$ is a distance on density operators (e.g., the trace distance). Operationally, this distance is evaluated by maximizing the distinguishability of the output states. In the special case $\epsilon = 0$, this reduces to the non-Markovianity with respect to the set $\mathcal{S}^0$ of Markovian-like maps. This formalism distinguishes the structural property of being exactly or approximately Markovian from the quantitative extent to which a given map departs from such behavior. In this way, $\epsilon$-Markovianity provides a hierarchy of increasingly relaxed notions of memorylessness. Ref.~\cite{Das_2021} further bounds $\mathcal{N}_\epsilon$ from above by a distance-based entanglement of the system–environment state generated by a dilation, plus the diameter of the set of separable states.

\subsection{Maximal non-Markovianity}
Beyond binary classifications of Markovian and non-Markovian dynamics, it is natural to ask whether different non-Markovian processes can be ordered according to the strength of their memory effects, which is particularly important when comparing dynamics that violate CP divisibility in qualitatively different ways. To understand this, consider an open quantum system $S$ whose evolution is described by a family of CPTP maps $\{\Lambda_t\}_{t\ge 0}$, such that $\rho(t)=\Lambda_t[\rho(0)]$. The dynamics is defined to be Markovian if it is CP divisible~\cite{Rivas_2010}; its violation signals non-Markovian behavior. To refine this binary classification, the notions of $k$-divisibility and maximally non-Markovian dynamics were introduced in~\cite{Dariusz2014} and are defined as follows.
\begin{definition} A quantum dynamical map $\{\Lambda_t\}_{t\ge 0}$ acting on a $d$-dimensional system is said to be maximally non-Markovian if it is not even positive divisible.
\end{definition}
In other words, there exist times $t\ge s$ for which no positive, trace-preserving intermediate map $V_{t,s}$ satisfies
\begin{align}
    \Lambda_t = V_{t,s}\circ \Lambda_s ,
\end{align}
and a dynamical map is said to be $k$-divisible if the intermediate propagator $V_{t,s}$ is $k$-positive for all $t \ge s$. Complete positivity corresponds to $k=d$, where $d=\dim(\mathcal{H}_S)$, while only positivity corresponds to $k=1$. This construction leads to a strict hierarchy of dynamical maps,
\begin{align}
\mathcal{D}_d \subset \mathcal{D}_{d-1} \subset \cdots \subset \mathcal{D}_1 ,
\end{align}
where $\mathcal{D}_k$ denotes the set of $k$-divisible maps. Based on this hierarchy, a non-Markovianity degree (NMD) is defined to quantify how far a given dynamics departs from CP divisibility. For a system of dimension $d$, the NMDs are given by
\begin{align}
\mathrm{NMD}(\Lambda_t) = d - k,
\end{align}
where $k$ is the largest integer for which the dynamics is $k$-divisible. Markovian dynamics correspond to $\mathrm{NMD}=0$, while dynamics that are not even positive divisible ($k=0$ by convention) attain the maximal value $\mathrm{NMD}=d$, and are therefore termed maximally non-Markovian. Such dynamics exhibit the strongest possible memory effects allowed by the system dimension. The structure of this classification closely mirrors the Schmidt number hierarchy in entanglement theory, where states are ordered according to the minimal Schmidt rank in their decompositions. This analogy enables the interpretation of maximally non-Markovian dynamics in direct correspondence with maximally entangled states, and motivates the introduction of non-Markovianity witnesses and monotones.

\section{$P$-divisibility vs information backflow}\label{sec:Comparison}
Divisibility-based and information-backflow-based approaches to quantum non-Markovianity are often discussed in tandem, yet they probe distinct structural and operational aspects of open quantum dynamics. While divisibility characterizes the intrinsic properties of the dynamical map itself, information backflow is defined by the behavior of suitable contractive quantities under evolution. The precise relationship between these notions has been clarified in Refs.~\cite{Andrzej2011,Erling_2018,Bylicka_2017} (see also Ref.~\cite{Wissmann2015}), and we summarize their key results here without repeating the general contractive framework already discussed in Section~\ref{subsec:Generalized contractive}.

Consider a family of CPTP maps $\{\Lambda_t\}_{t\ge 0}$ describing the reduced dynamics of a quantum system. Divisibility-based characterizations are formulated in terms of intermediate propagators. The evolution is said to be CP divisible if, for all $t \ge s \ge 0$, there exists a CPTP map $V_{t,s}$ as defined in Eq.~\eqref{eq:CP-div}.
Relaxing complete positivity leads to $ P$-divisibility, where the intermediate map $V_{t,s}$ is required to be positive and trace-preserving, but not necessarily completely positive.

Information backflow, on the other hand, is an operational notion identified through the temporary violation of monotonicity of contractive quantities under the dynamics. As discussed earlier (see Sec.~\ref{subsec:Generalized contractive}), contractive quantities cannot increase under CPTP maps, and any such increase signals a departure from memoryless behavior. Importantly, information backflow does not define a structural property of the map, but rather a dynamical feature revealed through suitable probes. It is therefore essential to specify the probe: below, ``information backflow'' refers to quantities defined on the system alone (system-only distinguishability, as in the BLP approach), unless an ancilla is explicitly included.

The logical relation between divisibility and information backflow can be made precise. If the evolution is CP divisible, then all intermediate propagators are CPTP, and hence for any contractive quantity $f$,
\begin{align}
f\!\left(\Lambda_t[\rho],\,\Lambda_t[\sigma]\right)
&=
f\!\left(V_{t,s}[\Lambda_s\rho],\,V_{t,s}[\Lambda_s\sigma]\right) \nonumber\\
&\le
f\!\left(\Lambda_s[\rho],\,\Lambda_s[\sigma]\right),
\end{align}
which guarantees the absence of information backflow. Therefore, CP divisibility is a sufficient condition for the monotonic decay of all contractive quantities, including ancilla-assisted ones.

However, CP indivisibility is not sufficient to imply system-only information backflow. As explicitly demonstrated in Ref.~\cite{Andrzej2011}, also in the example of Appendix~\ref{exmpl_RHP-BLP}, there exist CP-indivisible dynamical maps for which all system-only distinguishability quantities remain monotonically decreasing for all pairs of states. (By contrast, an ancilla-assisted contractive quantity always reveals CP-indivisibility; see Sec.~\ref{subsec:Generalized contractive}.) This shows that system-only information backflow is not a universal consequence of CP indivisibility. The precise structural boundary for system-only information backflow is given by the positivity of the intermediate propagators. If a dynamical map is $P$-divisible, then every distinguishability quantity on the system that is contractive under positive trace-preserving maps, such as the trace distance, decays monotonically in time, ruling out system-only information backflow. Conversely, for invertible dynamics, if $P$ divisibility is violated, then there necessarily exists a pair of states and prior probabilities $p_1,p_2$ such that the generalized (Helstrom-weighted) trace distance increases~\cite{Andrzej2011,Wissmann2015,Bae_2016},
\begin{align}
\frac{d}{dt} \big\|\Lambda_t[p_1\rho-p_2\sigma]\big\|_1 > 0 ,
\end{align}
signaling information backflow. Hence, for invertible dynamics, system-only information backflow (in this generalized sense) occurs if and only if $P$ divisibility is violated; for non-invertible dynamics, additional conditions are required~\cite{Erling_2018}.

These results establish the implication structure
\begin{align*}
&{\small\text{CP divisibility}
\Longrightarrow
\text{$P$ divisibility}}\\
&\qquad{\small\Longrightarrow
\text{No system-only information backflow}
}
\end{align*}
while the reverse implications do not hold in general (the second one is reversed for invertible dynamics with the generalized trace distance); in particular,
\begin{align*}
    &{\small\text{CP divisibility}\centernot \Longleftarrow \text{No system-only information backflow}.}
\end{align*}
In particular, CP indivisibility defines a broader class of non-Markovian dynamics than the class detected by system-only information backflow.

This section has provided a comprehensive overview of the principal frameworks for characterizing quantum non-Markovianity. Starting from CP-divisibility and its refinements, we discussed structural definitions based on intermediate dynamical maps, operational approaches based on state distinguishability and correlation revivals, and information-theoretic formulations using mutual and conditional mutual information. We further examined generalized contractive frameworks, which unify many existing criteria, and the process-tensor formalism, which reveals the fundamentally multi-time nature of quantum memory. Approximate and maximally non-Markovian dynamics were introduced to quantify deviations from ideal memoryless behavior and to order non-Markovian processes according to the strength of their memory effects. Finally, the explicit comparison between divisibility and information backflow clarified that system-only information backflow is not equivalent to CP indivisibility, but is instead linked to the breakdown of $P$ divisibility (precisely so for invertible dynamics). Together, these approaches highlight that quantum non-Markovianity is a multifaceted concept that cannot be fully captured by any single criterion.

\section{Genuine vs non-genuine backflow of information} \label{sec:QC_QNM}
The approaches discussed in Sec.~\ref{Defs_Markov} are designed primarily to detect memory effects and do not, by themselves, distinguish the nature of the information backflow or memory responsible for non-Markovian behavior in dynamics. In particular, CP indivisibility, information backflow, and correlation revivals do not determine whether the observed information backflow originates from quantum resources, such as coherence or entanglement between systems and the environment, or can be reproduced by purely classical correlations. As a result, many studies investigate whether the observed backflow is quantum or can also arise from classical mechanisms (such as convex mixing) in open quantum dynamics~\cite{Milz_2020,Breuer_2018-mixing-induced}. This motivates a finer classification of non-Markovian dynamics based on the classical or quantum character of information backflow. While the existing literature has extensively reviewed quantum non-Markovianity from multiple perspectives~\cite{Bassano2016,LI20181,Rivas_2014,Shrikant2023, Milz_2021}, a systematic and unified treatment that organizes and compares these notions specifically in terms of their classical and quantum origins remains lacking.

In this section, we will discuss the classical and quantum nature of quantum non-Markovianity in detail. A striking example is provided by classical environments, where information revival can occur even when the environment remains unaffected by the system’s dynamics~\cite{Franco2012, Budini2018}. This shows that an observed revival does not necessarily signal the presence of quantum non-Markovianity. Motivated by this, several works investigate how to separate the classical and quantum contributions to information revival that lead to quantum non-Markovianity, using approaches based on correlations, conditional structures, and temporal features~\cite{Giarmatzi2021, Banacki2023, Backer2024, Milz_2020}. Despite these advances, the terms ``revivals'' and ``backflows'' are still often used interchangeably, even though they correspond to different operational notions~\cite{Buscemi_2025}.

\subsection{Distinguishability-based information backflow}
\label{subsec:Banacki}

The BLP framework associates non-Markovianity with revivals of the trace distance,
as discussed in Sec.~\ref{subsec:BLP_QNM}. However, a revival alone does not specify
the physical nature of the memory responsible for it: an increase in
distinguishability does not necessarily imply that the information returning to the
system was stored quantum mechanically. Banacki \emph{et al.}~\cite{Banacki2023}
made this precise by showing that BLP information backflow can arise in dynamics
that are convex mixtures of evolutions without any quantum backflow, so that the
memory reduces to classical ignorance about which evolution is applied.

The construction is defined relative to a fixed orthonormal basis $B$ of the system,
which plays the role of a classical reference. Two states $\rho_1,\rho_2$ are called
indistinguishable by measurements in $B$ if they have the same diagonal in $B$,
i.e., $\Delta(\rho_1-\rho_2)=0$, where $\Delta$ denotes the completely dephasing map
in $B$. Such states differ only in their coherences in $B$, so any information that
discriminates them is quantum with respect to this reference. A dynamical map
$\Lambda_t$ is called \emph{elementary} with respect to $B$ if
\begin{align}
\frac{d}{dt}\big\|\Lambda_t(\rho_1-\rho_2)\big\|_1\le 0
\quad\text{for all } \rho_1,\rho_2 \text{ with } \Delta(\rho_1-\rho_2)=0,
\label{eq:elementary}
\end{align}
that is, if the coherence-encoded part of the information never flows back. Since
elementarity with respect to a given basis is not preserved under convex combinations
in general, the authors allow different bases for different components. A dynamics
is said to have \emph{no quantum information backflow} (in the strong sense) if it
admits a time-independent convex decomposition
\begin{align}
\Lambda_t=\sum_i p_i\,\Lambda_t^{(i)},\qquad p_i\ge0,\quad \sum_i p_i=1,
\label{eq:Banacki_decomp}
\end{align}
in which each $\Lambda_t^{(i)}$ is elementary with respect to some basis $B_i$. The
index $i$ then plays the role of a classical hidden variable that is inaccessible
to the observer, in close analogy with the mixing-induced memory discussed in
Sec.~\ref{subsec:Breuer_mixing_operational}. Subclasses is obtained by restricting the components: in dynamics of type~II and
type~I, each $\Lambda_t^{(i)}$ is, up to a time-dependent unitary rotation, diagonal
or block-diagonal (with respect to populations and coherences in $B_i$), while in
the most classical case, type~0, each component is \emph{generalized classical},
\begin{align}
\Lambda_t^{(i)}=U_i(t)\,\Lambda_t^{\mathrm{cl},i}\,U_i(t)^\dagger ,
\end{align}
where $\Lambda_t^{\mathrm{cl},i}$ acts on the populations in a fixed basis as a
stochastic matrix and erases coherences. Such dynamics nevertheless display BLP
non-Markovianity. For example, a qubit depolarizing dynamics whose contraction
factor oscillates in time, and hence exhibits trace-distance revivals, can be written
as an equal mixture of three classical dynamics acting in the eigenbases of
$\sigma_x$, $\sigma_y$, and $\sigma_z$~\cite{Banacki2023}. In this case, the observed
backflow is entirely attributable to classical mixing and should not be regarded as
a quantum phenomenon. Accordingly, a dynamics is said to exhibit quantum information
backflow in the weak sense if it shows BLP backflow but is not of type~0, and in the
strong sense if it admits no decomposition of the form~\eqref{eq:Banacki_decomp}.

For qubits, the authors characterize elementary maps completely and derive an
operational witness of quantum information backflow in the weak sense. The idea is
analogous to entanglement witnesses, with separable states replaced by the convex
hull $\mathcal{C}$ of Choi states of \emph{classical channels}, i.e., channels that
measure the input in some orthonormal basis and prepare states of another basis
according to a stochastic matrix. At any fixed time, the Choi state of a type-0
dynamics lies in $\mathcal{C}$. A Hermitian operator $W$ with $\mathrm{Tr}(W\rho)\ge0$
for all $\rho\in\mathcal{C}$ is a witness, and $\mathrm{Tr}(W\rho_{\Lambda_t})<0$
certifies that $\Lambda_t$ is not a convex mixture of classical channels, where
$\rho_{\Lambda_t}$ is the normalized Choi state of $\Lambda_t$. Writing
$\rho_{\Lambda_t}$ in terms of the Bloch vector $\mathbf{s}$ of its output marginal
and its correlation matrix $T$, optimizing over all such witnesses yields a single
computable condition~\cite{Banacki2023}:
\begin{align}
X(\rho_{\Lambda_t}) := |\mathbf{s}| + \|T\|_1 \le 1
\quad \Longleftrightarrow \quad \rho_{\Lambda_t}\in\mathcal{C}.
\label{eq:Banacki_witness}
\end{align}
Hence, if $X(\rho_{\Lambda_t})>1$ at some time $t\in[t_1,t_2]$, the dynamics is not
of type~0 on that interval. If it also shows BLP backflow, this backflow cannot be
simulated by a random choice of classical evolutions. For example, the identity
channel gives $X=3$, whereas the depolarizing example above satisfies $X<1$
throughout, consistent with its classically simulable backflow. The criterion is
one-directional: $X\le1$ at each time is necessary but not sufficient for type~0,
which requires a single decomposition valid over the whole interval. The extension
to dimension $d>2$ remains open.

This analysis establishes that information backflow, as captured by trace-distance revivals, is not a sufficient signature of quantum memory. Nor is it necessary, since quantum memory can be present in dynamics without trace-distance revivals (see Sec.~\ref{subsec:GC_quantum_memory}). Distinguishing quantum from classical backflow therefore requires criteria beyond distinguishability alone, such as the witness~\eqref{eq:Banacki_witness} or the multi-time criteria discussed below.

\subsection{Mixing-induced classical memory effects}
\label{subsec:Breuer_mixing_operational}

Breuer \emph{et al.}~\cite{Breuer_2018-mixing-induced} analyzed the emergence of non-Markovian behavior under convex mixing of quantum dynamical maps and clarified its interpretation from an information-flow perspective. Their work addresses an apparent paradox: a convex mixture of Markovian dynamics can exhibit non-Markovianity according to standard information-backflow criteria. Rather than contradicting the information-flow picture, this phenomenon is shown to arise from a well-defined operational mechanism that cleanly separates accessible (classical) and inaccessible information.

The setting involves a family of quantum dynamical maps $\{\Lambda_t^{(i)}\}$ that describe the reduced dynamics of an open system $S$. A mixed dynamics is defined as
\begin{align}
\Phi_t = \sum_i q_i\, \Lambda_t^{(i)},
\end{align}
where $q_i \ge 0,\ \sum_i q_i = 1$, which is again a CPTP map. Although each $\Lambda_t^{(i)}$ may generate Markovian evolution, characterized by a monotonic decrease of state distinguishability, the mixed map $\Phi_t$ can display revivals of distinguishability and is therefore non-Markovian in the sense of information backflow.

To provide an operational interpretation, the authors construct an explicit microscopic dilation of the mixing procedure. The system $S$ is coupled not only to environments $E_i$ generating the individual maps $\Lambda_t^{(i)}$, but also to an additional ancilla $A$ that stores the classical label $i$ indicating which map is applied. The ancilla is prepared in the classical state
\begin{align}
\rho_A = \sum_i q_i \ket{i}\!\bra{i},
\end{align}
and the global dynamics correlate $S$ with $A$ without generating entanglement: the joint state is of the classical--quantum form $\sum_i q_i\ket{i}\!\bra{i}_A\otimes\Lambda^{(i)}_t(\rho_S)$, which has zero quantum discord with respect to measurements on $A$. Tracing out the ancilla yields the mixed map $\Phi_t$. This construction allows a precise decomposition of information into two operationally distinct contributions. The internal information is defined as the distinguishability accessible by measurements on the system alone,
\begin{align}
I_{\mathrm{int}}(t)
= \frac{1}{2}\bigl\| \Phi_t[\rho_S^{(1)} - \rho_S^{(2)}] \bigr\|_1 ,
\end{align}
while the total information corresponds to the distinguishability when the ancilla is also accessible. The difference between these two quantities defines the external information,
\begin{align}
I_{\mathrm{ext}}(t) = I_{\mathrm{tot}}(t) - I_{\mathrm{int}}(t),
\end{align}
which quantifies information stored in correlations between the system and the ancilla. A central result is that any revival of internal information,
\begin{align}
\frac{d}{dt} I_{\mathrm{int}}(t) > 0 ,
\end{align}
must be accompanied by a decrease in external information. Operationally, this corresponds to information flowing from degrees of freedom inaccessible to the observer back into the system. Importantly, the ancilla represents a purely classical memory: its correlations with the system are classical in the above sense. This analysis establishes that mixing-induced non-Markovianity is entirely compatible with an information-flow interpretation. However, the backflow observed in such scenarios is not quantum. It originates from classical uncertainty about which dynamical map has been applied and from classical correlations between the system and an inaccessible ancilla. The resulting information backflow is therefore classical in nature, even though it manifests as non-Markovian behavior according to trace-distance criteria.

From an operational perspective, this work demonstrates that information-backflow-based measures of non-Markovianity do not, by themselves, distinguish between classical and quantum memory effects. Mixing-induced non-Markovianity provides a paradigmatic example in Appendix~\ref{exam:mixing} (also see Fig.~\ref{fig:placeholder}), where memory effects arise without any quantum coherence or entanglement in the memory degrees of freedom. This insight reinforces the need for additional structural or multi-time criteria when classifying non-Markovian dynamics into classical and genuinely quantum regimes.

\begin{remark}
A related mechanism, in which a classical description emerges from a fully quantum model once part of it is left unobserved, was analyzed for open quantum field dynamics in Ref.~\cite{Tomizuka2026}. There, a quantum system $\psi$ and a second quantum degree of freedom $\tilde h$ are both coupled to an unobserved environment. Tracing out the environment yields reduced dynamics with nonlocal noise and dissipation kernels, and hence a non-Markovian evolution. Performing a Wigner transform with respect to $\tilde h$ only, one obtains an operator $\hat W(\tilde h,\tilde\pi)$ acting on the Hilbert space of $\psi$, and $\tilde h$ behaves as an effectively classical variable whenever
\begin{align}
\hat W(\tilde h,\tilde\pi)\succeq 0 \quad \forall\,(\tilde h,\tilde\pi),
\end{align}
so that $\hat W$ has the structure of a classical--quantum state. A sufficient condition for this positivity to be preserved by the evolution is the positivity of the kernels that couple the forward and backward branches of $\psi$ and that govern the noise of $\tilde h$. This is a condition on the process over the whole evolution interval, not on two-time propagators. In the Markovian (short-memory) limit, the model reproduces the classical–quantum dynamics of Oppenheim and collaborators, together with its decoherence--diffusion trade-off.. As in the mixing-induced example above, the apparently classical behavior arises from degrees of freedom that are inaccessible in the reduced description. We note that this criterion addresses the classicality of a subsystem rather than the classical or quantum character of the memory; connecting it to witnesses of genuine quantum non-Markovianity would be an interesting direction.
\end{remark}

\subsection{Dynamical-map-based witnesses of quantum memory}
\label{subsec:choi_quantum_memory}

A separation between classical and quantum contributions to non-Markovian dynamics was recently introduced in Ref.~\cite{Backer2024}. Unlike distinguishability or correlation-based approaches, this framework focuses directly on the type of memory required to reproduce a given non-Markovian dynamics using only local information about the system. The central question addressed is as follows: Given the reduced dynamics of a system at different times, can the observed memory effects be simulated using only classical memory, or do they necessarily require a persistent quantum memory?

The setting considers a discrete-time dynamics described by a sequence of CPTP maps
\begin{align}
\mathcal{E} = (\mathcal{E}_1, \mathcal{E}_2),
\end{align}
where $\mathcal{E}_1$ maps the system state from the initial time $t_0$ to an intermediate time $t_1$, and $\mathcal{E}_2$ maps the state from $t_0$ directly to a later time $t_2$. Importantly, $\mathcal{E}_2$ is not assumed to factorize as $\mathcal{E}_{2:1}\circ\mathcal{E}_1$, and the intermediate propagator $\mathcal{E}_{2:1}$ need not be CPTP. This allows the framework to capture generic non-Markovian behavior.

The notion of classical memory is defined operationally at the level of dynamical maps. The dynamics $\mathcal{E}$ is said to be realizable with classical memory if and only if there exists a Kraus decomposition of the first map,
\begin{align}
\mathcal{E}_1(\rho)
= \sum_i M_i \rho M_i^\dagger,
\end{align}
and a collection of CPTP maps $\{\Phi_i\}$ such that the second map can be written as
\begin{align}
\mathcal{E}_2(\rho)
= \sum_i \Phi_i\!\left(M_i \rho M_i^\dagger\right).
\label{eq:classical_memory_choi}
\end{align}
Equation~\eqref{eq:classical_memory_choi} has a clear interpretation: the first step corresponds to a generalized measurement with classical outcomes $i$, and the outcome label is stored in a classical register. The second step applies a conditional quantum channel depending only on this classical information. No quantum system needs to persist across time steps; hence, no quantum memory is required. If such a representation is impossible, the dynamics necessarily rely on a quantum memory. To detect when Eq.~\eqref{eq:classical_memory_choi} fails, the authors employ the Choi--Jamio{ł}kowski isomorphism. For a CPTP map $\mathcal{E}$ acting on a $d$-dimensional system, the associated Choi state is defined as
\begin{align}
C[\mathcal{E}]
= (\mathcal{E}_S\otimes \mathds{1}_A)
\bigl(\ket{\phi^+}\!\bra{\phi^+}_{SA}\bigr),
\end{align}
where $A$ is an ancilla system isomorphic to $S$ and $\ket{\phi^+}$ is the normalized maximally entangled state. The Choi state encodes all operational properties of the channel, and allows one to study the dynamics using tools from entanglement theory.

The key quantity introduced is the \emph{entanglement of assistance}. Given a bipartite state $\rho_{SA}$ and an entanglement measure $E(\cdot)$, the entanglement of assistance is defined as
\begin{align}
E_a(\rho_{SA})
= \max_{\{p_k,\ket{\psi_k}\}}
\sum_k p_k\, E(\ket{\psi_k}),
\end{align}
where the maximization is over all pure-state decompositions
$\rho_{SA}=\sum_k p_k \ket{\psi_k}\!\bra{\psi_k}$.
Operationally, $E_a$ quantifies the maximum average entanglement that can be generated by optimally resolving the state into pure components. It is to be contrasted with the entanglement of formation $E_f(\rho_{SA})=\min_{\{p_k,\ket{\psi_k}\}}\sum_k p_k E(\ket{\psi_k})$, the corresponding convex roof, which is convex and non-increasing under local operations.

Let $C_1=C[\mathcal{E}_1]$ and $C_2=C[\mathcal{E}_2]$ be the Choi states corresponding to the two dynamical maps. The central result of Ref.~\cite{Backer2024} is the following sufficient criterion:
\begin{align}
E_a(C_1) < E_f(C_2).
\label{eq:quantum_memory_witness}
\end{align}
If~\eqref{eq:quantum_memory_witness} is satisfied, then the dynamics require quantum memory. The logic behind this inequality is purely information-theoretic. If the dynamics admitted a classical-memory realization of the form~\eqref{eq:classical_memory_choi}, then $C_1$ would first be decomposed into pure components corresponding to the Kraus operators $M_i$. Subsequent conditional CPTP maps $\Phi_i$ could only reduce entanglement on average. As a result, the entanglement of formation present in $C_2$ would be upper bounded by the entanglement of assistance of $C_1$, i.e., $E_f(C_2)\le\sum_i p_i E(\psi_i)\le E_a(C_1)$, where $p_i\ket{\psi_i}\!\bra{\psi_i}=(M_i\otimes\mathds{1})\ket{\phi^+}\!\bra{\phi^+}(M_i\otimes\mathds{1})^\dagger$. Satisfaction of inequality~\eqref{eq:quantum_memory_witness} therefore rules out all classical-memory models. This criterion has two immediate implications for the classification of quantum non-Markovianity.
First, the observation of non-Markovianity through revivals of distinguishability or correlations alone is not sufficient to certify the presence of quantum memory.
Second, the criterion defines a quantum-memory witness that relies only on the dynamical maps $\mathcal{E}_1$ and $\mathcal{E}_2$ from the initial time (no intermediate interventions) and does not require access to multi-time statistics (such as process tensors) or the environment. It establishes a direct connection between quantum non-Markovianity and the impossibility of constructing a measure-and-prepare type representation such as Eq.~\eqref{eq:classical_memory_choi}. Whenever the criterion~\eqref{eq:quantum_memory_witness} is satisfied, no classical records can reproduce the dynamics, and the memory stored in the environment must be quantum. Memory effects in non-Markovian dynamics are also investigated on IBM superconducting quantum processors using a collision-model circuit, demonstrating that current noisy hardware can verify quantum memory in single-qubit dynamics and witness it in two-qubit systems through a toy model~\cite{backer2025}.

The criterion based on the entanglement of assistance provides a general method for detecting the necessity of quantum memory. However, evaluating entanglement measures appearing in Eq.~\eqref{eq:quantum_memory_witness} becomes computationally demanding for systems beyond qubits.
To address this limitation, Ref.~\cite{Beyer_2025} reformulated the witness in a more tractable form using entropic bounds. The construction again considers an ancilla $A$ that remains unaffected by the system dynamics. Starting from an initial joint state $\rho^{SA}_0$, the system evolves under the dynamical maps $\mathcal{E}_{t}$, leading to
\begin{align}
\rho^{SA}_t=(\mathcal{E}_t\otimes\mathds{1}_A)(\rho^{SA}_0).
\end{align}
In the entanglement-based formulation discussed above, the states $\rho^{SA}_{t_1}$ and $\rho^{SA}_{t_2}$ correspond to the Choi states of the dynamical maps when $\rho^{SA}_0$ is maximally entangled. Using bounds that relate the entanglement of assistance and the entanglement of formation to von Neumann entropies($E_a\le S_S$ and $E_f\ge\max\{-S_{S|A},-S_{A|S}\}$ for the entropy of entanglement), the witness can be expressed in terms of subsystem and conditional entropies. This leads to the sufficient condition
\begin{align}
S_S(\rho^{SA}_{t_1}) < \max\!\left\{-S_{S|A}(\rho^{SA}_{t_2}),
             -S_{A|S}(\rho^{SA}_{t_2})\right\},
\label{eq:entropic_memory_witness}
\end{align}
where $S_S$ denotes the entropy of the reduced system state and $S_{S|A}$ and $S_{A|S}$ are the corresponding conditional entropies. Satisfaction of Eq.~\eqref{eq:entropic_memory_witness} rules out any realization of the dynamics using classical memory in the sense of Eq.~\eqref{eq:classical_memory_choi}. Importantly, this entropic formulation depends only on reduced system dynamics and entropy evaluations, making it applicable to higher-dimensional systems and continuous-variable dynamics where direct computation of entanglement measures is typically infeasible~\cite{Beyer_2025}.

\subsection{Operational characterization of classical processes}
\label{subsec:Milz_operational}

Milz \emph{et al.}~\cite{Milz_2020} developed an operational framework that determines when the memory of a non-Markovian process can be described classically, directly at the level of experimentally accessible multi-time statistics. The central idea is to characterize the nature of memory not through divisibility or correlation revivals, but through the classical simulability of sequential measurement outcomes obtained from an open quantum process.

Consider a system $S$ interacting with an environment $E$, undergoing a global unitary evolution. At a sequence of times $t_1<\cdots<t_n$, projective measurements $\{P_x\}$ are performed on the system only. The observed joint probability distribution of outcomes $\vec{x}=(x_1,\ldots,x_n)$ is given by
\begin{align}
&P(x_n,\ldots,x_1)
=
\Tr\!\left[
\mathcal{P}_{x_n}\circ
\mathcal{U}_{t_n:t_{n-1}}
\circ \cdots \circ
\mathcal{P}_{x_1}\circ
\mathcal{U}_{t_1:t_0}
(\rho_{SE})
\right],
\end{align}
where $\mathcal{U}_{t_{k+1}:t_k}$ denotes the global system--environment unitary evolution between successive times and $\mathcal{P}_x(\cdot)=(P_x\otimes\mathds{1}_E)(\cdot)(P_x\otimes\mathds{1}_E)$ acts on the system only. (Writing the statistics as a composition of reduced system maps would presuppose divisibility, i.e., the absence of memory.) These probabilities fully characterize the experimentally accessible temporal statistics of the process.

The process is said to be classical (with respect to the chosen measurement basis) if, for all choices of times and outcomes, the joint probabilities are Kolmogorov consistent, i.e., summing over the outcome at any time $t_k$ reproduces the statistics obtained when no measurement is performed at $t_k$,
\begin{align}
\sum_{x_k}P(x_n,\ldots,x_k,\ldots,x_1)=P(x_n,\ldots,\cancel{x_k},\ldots,x_1)\quad\forall k .
\end{align}
(Kolmogorov consistency should not be confused with the Chapman--Kolmogorov equation~\eqref{eq:CK_eq}.) By the Kolmogorov extension theorem, such statistics can be described by an underlying classical stochastic process, for example of the form
\begin{align}
P(x_n,\ldots,x_1)=\sum_{\lambda_n,\ldots,\lambda_1}\prod_{k=1}^{n}P(x_k|\lambda_k)\,
P(\lambda_n|\lambda_{n-1})\cdots P(\lambda_2|\lambda_1)P(\lambda_1),
\end{align}
which corresponds to a classical process with memory. In this case, any apparent information backflow can be fully simulated by classical correlations stored in the environment, even if the reduced dynamics is non-Markovian.

Using the process-tensor (quantum-comb) formalism, Milz \emph{et al.} related this operational notion of classicality to a structural constraint on the process tensor $T_{n:0}$. In particular, a process yields classical statistics if $T_{n:0}$ is diagonal in the tensor-product basis induced by the measurement projectors,
\begin{align}
T_{n:0}
=
\sum_{\vec{x},\vec{x}'}
p(\vec{x},\vec{x}')
\bigotimes_{k=0}^{n-1}
\ket{x_k}\!\bra{x_k}
\otimes
\ket{x_k'}\!\bra{x_k'} .
\end{align}
This diagonal structure ensures that all temporal correlations can be encoded in classical random variables, ruling out any quantum contribution to information backflow. The necessary and sufficient condition is that measurements in this basis cannot distinguish identity from completely dephasing interventions at any subset of times~\cite{Milz_2020}.

A key insight of this work is that, in the presence of memory, the absence of
detectable coherence on the system is no longer sufficient for classicality: there
are processes in which the system state never displays coherences, yet the
statistics are nonclassical. The relevant resource is instead the basis-dependent
quantum discord between system and environment. A non-Markovian process is
classical if and only if it can be modeled by an initial system--environment state with vanishing basis-dependent discord and by system--environment maps that are non-discord-generating-and-detecting (NDGD)~\cite{Milz_2020}. Such maps may create discord, but not in a way that later measurements on the system can detect. Vanishing discord at all times is sufficient but not necessary for classicality, and the discord must vanish with respect to the measurement basis.

The framework further identifies a class of quantum non-Markovian processes: processes for which no choice of measurement basis yields classically simulable multi-time statistics. Such behavior is impossible for Markovian processes and cannot be captured by two-time criteria such as CP divisibility or trace-distance revivals. Instead, it is intrinsically linked to the multi-time quantum correlations encoded in the process tensor.

Overall, this operational approach shows that the classical or quantum nature of information backflow is determined neither by divisibility nor by correlation revivals alone, but by the impossibility of reproducing sequential measurement statistics using a classical stochastic model. This places the distinction between classical and quantum non-Markovianity firmly at the level of multi-time operational data rather than reduced dynamical maps.

\subsection{Process-tensor structure of quantum memory}
\label{subsec:GC_quantum_memory}

A process-tensor framework for distinguishing between classical and quantum information backflow, along with an operational approach, was introduced by Giarmatzi and Costa~\cite{Giarmatzi2021}. Rather than relying on two-time dynamical maps or information-backflow measures, this approach characterizes the nature of memory directly from the multi-time structure of the process tensor $T_{n:0}$ itself, as discussed in Section~\ref{subsec:process_QNM}. While non-Markovianity is identified by the failure of factorization of $T_{n:0}$ into independent one-step channels, the type of memory is determined by the separability properties of $T_{n:0}$ across temporal partitions.

Processes with classical memory, i.e., those realizable by measuring the environment and feeding forward classical outcomes, admit a convex (separable) decomposition of the form
\begin{align}
T_{n:0}^{\mathrm{cl}}
=\sum_\lambda p_\lambda
\bigotimes_{k=0}^{n-1}
T_{k+1:k}^{(\lambda)},
\label{eq:GC_classical_memory}
\end{align}
where $p_\lambda \ge 0$, $\sum_\lambda p_\lambda = 1$, and each $T_{k+1:k}^{(\lambda)}$ is a positive operator corresponding to a (not necessarily normalized) completely positive map between consecutive times. Eq.~\eqref{eq:GC_classical_memory} captures the fact that all temporal correlations can be mediated by a classical hidden variable $\lambda$ stored in the environment and propagated forward in time. Processes with classical memory, by definition, admit a measurement-and-feedback realization and do not require any coherent quantum degree of freedom to persist across time steps.

If no decomposition of the form~\eqref{eq:GC_classical_memory} exists, the process tensor is entangled across time, which certifies that the process has quantum memory: the environment must preserve quantum coherence between different system–environment interactions. Temporal entanglement of the process tensor thus provides a precise and structural signature of quantum information backflow.

This distinction enables a direct connection between entanglement theory and the detection of quantum memory. Since $T_{n:0}$ is a positive operator, standard entanglement witnesses can be employed as quantum-memory witnesses. A Hermitian operator $W$ is such a witness if
\begin{align}
\Tr\!\left(W T_{n:0}^{\mathrm{cl}}\right) \ge 0
\end{align}
for all classical-memory processes, while
\begin{align}
\Tr\!\left(W T_{n:0}\right) < 0
\end{align}
certifies the presence of quantum memory. Importantly, evaluating these witnesses does not require full-process tomography; the witnesses can be decomposed into experimentally accessible operations at each time step.

A central implication of this analysis is that neither CP divisibility nor information backflow based on distinguishability is sufficient to characterize quantum memory. Processes with quantum memory may generate divisible dynamical maps, while processes with purely classical memory may violate divisibility and exhibit revivals of distinguishability~\cite{Giarmatzi2021}. The classical and quantum distinction is therefore invisible at the level of two-time maps and emerges only from the entanglement structure of the multi-time process tensor. This provides a rigorous and conceptually clean certification of quantum non-Markovianity: quantum information backflow is certified whenever the process tensor is temporally entangled (the converse need not hold, since a separable decomposition with non-normalized components need not be realizable with classical memory).

We have discussed different approaches in this section on why the common signature of non-Markovian behavior, such as the indivisibility of CP maps or revivals of distinguishability, is not sufficient to distinguish between quantum and classical information backflows. This clearly establishes that information revivals can also be explained classically in some cases, particularly in the convex mixing of Markovian processes (cf. Appendix~\ref{exam:mixing}), where the memory is stored in a classical form in an inaccessible ancilla, rather than quantum mechanically, either as coherence or entanglement. Various different paradigms have also been presented on how this distinction can be formally and precisely established: based on distinguishability witnesses that exclude classical hidden-variable models, based on indicators derived from the entanglement properties of Choi matrices of the dynamical maps, and on process tensor and sequential measurement statistics-based fully operational multi-time indicators of quantum behavior. One thing that emerges clearly in these approaches is that information backflow is neither necessary nor sufficient for quantum memory: genuine quantum memory is certified only when no classical-memory model can reproduce the observed data, i.e., when the process demands quantum correlations across time.

\subsection{Non-causal information revivals}\label{sec:non_causal_backflow}

The work of Ref.~\cite{Buscemi_2025} introduces a new conceptual distinction in the study of quantum non-Markovian dynamics by separating the notions of \emph{information revival} and \emph{information backflow}. While these two notions are often treated as equivalent in the literature, the central insight of this work is that an increase in correlations in time does not necessarily imply that information flows back from the environment to the system. The starting point is the operational observation that non-Markovianity is commonly witnessed through violations of data-processing inequalities, such as revivals of the quantum mutual information between a system and a reference. These revivals are fully determined by the reduced system dynamics and are therefore operationally accessible. However, their interpretation depends on the underlying system--environment model, which is not directly observable. The key conceptual shift introduced in this paper is the identification of a class of information revivals, termed \emph{noncausal revivals}, that can be explained without invoking any flow of information from the environment back to the system. Such revivals arise because part of the relevant information is stored in degrees of freedom that never interact with the system and are causally disconnected from its dynamics. When these hidden degrees of freedom are taken into account, the apparent violation of data-processing inequalities disappears. This idea is formalized by introducing inert extensions of the dynamics. A revival is defined to be noncausal if there exists an extension, remaining completely inert throughout the evolution, such that the revival vanishes when correlations with this extension are taken into account. Importantly, the existence of such an extension depends only on the intermediate tripartite configuration and not on the details of the later dynamics.

A central result of the paper is that noncausal revivals are exactly characterized by a conditional mutual information condition. Specifically, all possible revivals are noncausal if and only if the intermediate state satisfies
\begin{align}
I(R;E' \mid Q'F)=0
\end{align}
for some inert extension $F$, where $R$ is the reference, and $Q'$ and $E'$ denote the system and the environment at the intermediate time (see Ref.~\cite{Buscemi_2025} for the precise setting). This condition is directly linked to the notion of squashed quantum non-Markovianity introduced in Sec.~\ref{subsec:sQNM} at the level of states. In this sense, squashed non-Markovianity quantifies how far a given intermediate configuration is from allowing a purely noncausal explanation of future revivals. This connection offers a new interpretation of squashed quantum non-Markovianity, measuring the minimal amount of genuinely quantum memory that cannot be removed by conditioning on additional, causally disconnected systems. Small squashed non-Markovianity guarantees that any revival can be approximately explained as noncausal, while a large value implies the unavoidable presence of genuine backflow. The paper further introduces an operational, system-only criterion that is sufficient to certify genuine information backflow, without any reference to the environment. This criterion is formulated in terms of squashed entanglement between the system and the reference at the final time, showing that certain revivals cannot be explained by noncausal mechanisms alone.

A major conceptual consequence of this refined viewpoint is the resolution of the long-standing non-convexity problem of quantum Markovianity. While convex mixtures of Markovian processes may display information revivals, the paper shows that such revivals are always noncausal. As a result, when genuine backflows are isolated by excluding noncausal revivals, the resulting notion of non-Markovianity becomes convex. This observation opens the door to a consistent resource-theoretic formulation of genuine quantum non-Markovianity at the level of processes. Overall, this work offers a new perspective on non-Markovian dynamics: information revivals are treated as operational signatures, while genuine quantum non-Markovianity is identified with those revivals that cannot be explained without the causal exchange of information with the environment. By connecting process-level non-Markovianity with state-based squashed non-Markovianity, the paper offers a unified and physically transparent understanding of memory effects in open quantum systems.

\subsection{Quantum switch-induced non-Markovianity}

A natural setting, in which one can ask whether observed backflow of information is genuinely quantum, is provided by the quantum switch, which coherently superposes the order of two channels using a control qubit. Anand \emph{et al.}~\cite{Anand2025} studied the switch from a dynamical viewpoint and identified a form of quantum memory that emerges from its action. It was examined for depolarizing dynamics, including the case of a noisy control qubit and a noisy final measurement on it. By deriving a Lindblad-type description of the reduced action of the switch, they showed that the induced memory gives rise to non-Markovianity, even though the constituent channels are Markovian. This ``emergent non-Markovianity'' can be attributed to the switch itself by comparison with standard measures of non-Markovianity, so the switch can be regarded as a source of non-Markovian quantum memory. The measures employed there were designed to detect non-Markovianity rather than to identify the origin of the memory, which naturally raises the question of whether this emergent memory is genuinely quantum. This is the question addressed below.

The criteria discussed above can also be used to ask whether the memory effects generated by a causal-order resource are genuinely quantum. For a pair of channels $\mathcal{N}_1$ and $\mathcal{N}_2$ with Kraus operators $\{K^{(1)}_i\}$ and $\{K^{(2)}_j\}$, the Kraus operators of the quantum switch, the canonical realization of indefinite causal order, are
\begin{align}
S_{ij} = K^{(2)}_j K^{(1)}_i \otimes |0\rangle\langle 0|_c + K^{(1)}_i K^{(2)}_j \otimes |1\rangle\langle 1|_c ,
\label{eq:switch_kraus}
\end{align}
and the control is measured in the coherent basis $\{|\pm\rangle\}$ after the evolution, with $|\pm\rangle=(|0\rangle\pm|1\rangle)/\sqrt{2}$. Although the quantum switch is known to provide operational advantages in communication, computation, metrology, and thermodynamics, the resource responsible for them is not settled. In particular, it has been shown that placing individually Markovian channels in a switch can produce non-Markovian information backflow, so memory has been proposed as a possible source of these advantages. This raises the question of whether such switch-induced memory is genuinely quantum.

In~\cite{GangwarMaity2026}, the authors addressed this question by combining the LFS criterion of Sec. 3.3 with the non-causal revival framework of Sec. 5.6. A reference R, initially correlated with Q, evolves trivially. After the switch acts and the control is post-selected on |+⟩, revivals of I(R;Q) are detected. A revival is classified as non-causal if some inert extension F, never acted upon, makes I(R;QF) monotonic.

In a discrete-time example, a Bell state RQ is used with completely depolarizing and X-dephasing channels generated by unitary interactions with a maximally mixed and a classically correlated environment. There, the constituent processes already show non-causal revivals. The switch inherits them: the same revival appears in fixed order, and purifying the environments makes I(R;QF) constant.

In a second example, two one-parameter families of CP-divisible channels are inserted into the switch, with a Werner state RQ. The channels are generalized amplitude damping at infinite temperature, acting in the computational and the {|±⟩} basis, and t labels their strength rather than a time within a single run. The switch produces a small revival of I(R;Q) that is absent in each channel, in either fixed order, and in their incoherent mixture; this is the emergent non-Markovianity of~\cite{Anand2025}. Because the probability of the outcome |+⟩ is independent of the input, each snapshot is obtained from the initial state by a CPTP map. The purifying extension does not remove this revival, but a classical inert extension recording in which of the states |±⟩ one environment was prepared does.

Hence, for the standard switch, with each channel acting on its own environment, the switch-induced revivals are non-causal and do not certify genuine quantum non-Markovianity. Coherently ordering the two interactions with a shared environment is a different process. In the continuous example, revivals then arise already in fixed order from environmental reuse, and under coherent control none of the inert extensions tested removes them completely, so their classification remains open.

Conceptually, this extends the lessons of Secs.~\ref{subsec:Breuer_mixing_operational} and~\ref{sec:non_causal_backflow} from convex mixing to coherent control of causal order. Even a coherent superposition of causal orders can produce backflow-like signatures that, according to the squashed/non-causal criterion, come from an incomplete description of the environment rather than from a genuinely quantum non-Markovian effect. The result also bears on the origin of switch advantages: at least in the examples considered, these advantages do not appear to stem from genuine quantum non-Markovian memory, which points to other resources, such as control coherence or control dimension, as their possible source. The analysis is limited to the examples studied and to a state-based (two- and three-time) diagnostic, so a multi-time analysis based on process tensors (Secs.~\ref{subsec:Milz_operational} and~\ref{subsec:GC_quantum_memory}) would be a natural next test.

\subsection{Genuine quantum non-Markovianity in quantum states}\label{subsec:sQNM}
The work~\cite{Gangwar2024} introduces a conceptually different way of looking at quantum non-Markovianity by shifting the focus from dynamics to quantum states and by explicitly separating genuine quantum memory from non-genuine (classical or mixing-induced) contributions. Instead of asking whether a process shows revivals of distinguishability or violates divisibility, the central question is: which part of the observed non-Markovianity is intrinsically quantum and which part can be removed by classical conditioning?

We start from the observation that for a tripartite quantum state $\rho_{ABC}$, quantum non-Markovianity is traditionally characterized by the quantum conditional mutual information (QCMI)~\cite{Hayden_2004},
\begin{align}
I(A;C|B)_\rho
= S(AB)_\rho + S(BC)_\rho - S(B)_\rho - S(ABC)_\rho .
\end{align}
While $I(A;C|B)=0$ characterizes quantum Markov states, a nonzero QCMI does not necessarily imply a genuinely quantum non-Markovian state, since classical probabilistic mixtures of Markov states can also yield $I(A;C|B)>0$. This reveals that QCMI detects non-Markovianity, but not its nature of origin. To resolve this, in~\cite{Gangwar2024}, a \emph{squashed quantum non-Markovianity} (sQNM) measure has been introduced, which removes all non-genuine contributions by allowing extensions of the conditioning system. It is defined as
\begin{align}
N_{\mathrm{sq}}(A;C|B)_\rho
:= \frac{1}{2}\inf_{\rho_{ABCE}} I(A;C|BE)_\rho ,
\end{align}
where the infimum is taken over all extensions $\rho_{ABCE}$ such that
$\Tr_E[\rho_{ABCE}] = \rho_{ABC}$. The key insight is that any non-Markovianity that disappears after conditioning on an extended system $BE$ cannot be genuinely quantum. Only the part that survives this minimization corresponds to an intrinsic quantum non-Markov state. This construction leads to a clear conceptual separation:
\begin{itemize}
\item \emph{Non-genuine quantum non-Markovianity}, arising from classical mixing or conditioning, which can be removed by extending the conditioning system.
\item \emph{Genuine quantum non-Markovianity}, quantified by sQNM, which cannot be eliminated by any such extension.
\end{itemize}

A major new insight of this approach is that genuine quantum non-Markovianity forms a proper quantum resource. The set of states with vanishing sQNM is convex, reflecting the fact that classical mixing cannot create genuine quantum non-Markov states. Moreover, sQNM satisfies fundamental information-theoretic properties such as convexity, monogamy, additivity on tensor-product states, and asymptotic continuity. These properties are not simultaneously satisfied by QCMI itself or by earlier state-based notions of non-Markovianity. This framework also clarifies the relation between non-Markovianity and entanglement. For pure tripartite states $\ket{\psi}_{ABC}$, sQNM reduces to
\begin{align}
N_{\mathrm{sq}}(A;C|B)_\psi
= \tfrac{1}{2} I(A;C)_\psi ,
\end{align}
showing that genuine quantum non-Markovianity is directly tied to bipartite quantum correlations between the non-conditioning systems. More generally, sQNM is lower bounded by the squashed entanglement of $\rho_{AC}$ and constrained by extendibility, highlighting the monogamous character of genuine quantum non-Markovianity.

Overall, this paper introduces a new perspective on non-Markovianity: instead of identifying memory through dynamical signatures alone, it isolates the genuinely quantum part of memory at the level of states using conditional correlations and extensions. This viewpoint provides a clean, operationally meaningful foundation for distinguishing which non-Markovian effects are truly quantum from those that are artifacts of classical conditioning, and it opens the way to a unified treatment of genuine quantum non-Markovianity in both states and processes.

\section{Temporal correlation-based non-Markovianity}
\label{sec:temporal_corr}

The frameworks of Sec.~\ref{sec:QC_QNM} ask whether a process is Markovian and, if not, what kind of memory is responsible. A complementary line of work instead \emph{quantifies the temporal correlations themselves}: how much correlation a process transmits through the system, how much it carries through the environment, and how these amounts constrain one another. We review two such approaches: a witness built on temporal steering and a single-operator description of a dynamics as a ``state over time''. These approaches are primarily designed to quantify temporal correlations, and the separation of classical from quantum memory, in the sense of Sec.~\ref{sec:QC_QNM}, is addressed to varying extents; we comment on this where relevant.

\subsection{Temporal steering-based non-Markovianity}
\label{sec:temporal_steering}

The correlation-based quantifiers of Sec.~\ref{sec:correlation_based} rely on an ancilla. A complementary, ancilla-free route uses \emph{temporal
steering}~\cite{Chen2016}. Alice performs a measurement $\{M_{a|x}\}$ (with
$F_{a|x}=M_{a|x}^\dagger M_{a|x}$, $\sum_a F_{a|x}=\mathds{1}$) on a single
system in state $\rho_0$, so that
$\rho_0\mapsto\rho_{a|x}=M_{a|x}\rho_0 M_{a|x}^\dagger/p(a|x)$.  The post-measurement state is sent through the dynamics $\Lambda_t$, and Bob performs state tomography to obtain the (temporal) assemblage $\{\sigma^{T}_{a|x}\}_{a,x}$, i.e., the collection of subnormalized conditional states $\sigma^{T}_{a|x}=p(a|x)\,\Lambda_t(\rho_{a|x})$; for each setting $x$, it encodes the ensemble $ \{p(a|x),\Lambda_t(\rho_{a|x})\}$~\cite{Pusey2013,Cavalcanti_2017, Skrzypczyk2014,Chen2016}. The assemblage is \emph{unsteerable} if it admits a hidden-state form,
\begin{align}
\sigma^{T,\mathrm{US}}_{a|x}=\sum_\lambda P(\lambda)P(a|x,\lambda)\,\sigma_\lambda ,
\end{align}
meaning that it can be reproduced by a hidden-state model, without any genuine steering by Alice's measurement. This happens either because the dynamics washes out her influence or because her
results could have been fabricated by a classical strategy.

\begin{definition}[Temporal steerable weight~\cite{Chen2016}]
Decompose $\sigma^{T}_{a|x}=\mu\,\sigma^{T,\mathrm{US}}_{a|x}+(1-\mu)\,\sigma^{T,S}_{a|x}$
and let $\mu^{*}$ be the largest admissible $\mu$. The temporal steerable
weight is $\mathrm{TSW}=1-\mu^{*}$, where
\begin{align}
  &\mu^{*}=\max \operatorname{Tr}\sum_\lambda\tilde\sigma_\lambda \nonumber \\
  &
  \sigma^{T}_{a|x}-\sum_\lambda D_\lambda(a|x)\,\tilde\sigma_\lambda\ge 0\ \ \forall a,x,
  \qquad \tilde\sigma_\lambda\ge 0,
\end{align}
with $D_\lambda(a|x)$ the extremal deterministic response functions. This is a
semidefinite program~\cite{Stephen_1996,Pusey2013,Skrzypczyk2014}.
\end{definition}

The key property is that $\mathrm{TSW}$ is non-increasing under CPTP maps acting on Bob's side. Together with divisibility, it therefore decays monotonically under CP-divisible dynamics,
and a revival signals memory. As in the BLP construction, a quantifier is obtained
by integrating the positive slope,
\begin{align}
\mathcal{N}_{\mathrm{TSW}}=\int_{\dot{\mathrm{TSW}}(t)>0}\frac{d}{dt}\mathrm{TSW}\big[\Lambda_t,\rho_0\big] dt.
\end{align}

Because $\mathrm{TSW}$ vanishes for ``weak'' temporal correlations,
$\mathcal{N}_{\mathrm{TSW}}$ is a witness of \emph{strong} non-Markovianity, and it
is sufficient but not necessary. In a spin-boson model solved with the hierarchical equations of motion, Ref.~\cite{Chen2016} compares it with the
entanglement-based RHP measure~\cite{Rivas_2010}. Both behave similarly as a
function of bath cutoff frequency and temperature, but the entanglement-based
quantity detects memory over a wider range of system--bath coupling, consistent
with the hierarchy (Bell-nonlocal) $\subset$ (steerable) $\subset$ (entangled).
The practical advantage of steering is that only an assemblage is required, with no
ancilla and no process tomography. TSW refers to a hidden state (classical-realism) model for the \emph{assemblage}, which is conceptually
distinct from the classical-memory models of Sec.~\ref{sec:QC_QNM}. Whether TSW revivals can also certify genuinely quantum memory is an interesting open question.

\subsection{States over time and pseudo-density operators}
\label{sec:state_over_time}

The process tensor of Sec.~\ref{subsec:process_QNM} encodes the response of a process to arbitrary interventions.
A different, non-interventional description assigns to the evolution of a system at times $t_0<\dots<t_n$ a single operator on $\mathcal{H}_{S_0}\otimes\cdots\otimes\mathcal{H}_{S_n}$, called a \emph{state over time}, in direct analogy with the joint distribution $P(x_n|x_{n-1})\cdots P(x_1|x_0)P(x_0)$ that Eq.~\eqref{classical_markov} assigns to a classical Markov chain.
Here $S_k$ denotes the system at time $t_k$, with $\mathcal{H}_{S_k}$ of dimension $d_k$ and a fixed orthonormal basis $\{\ket{i}\}$.
We follow Refs.~\cite{Fullwood2022,Fullwood2025}, working in the Schr\"odinger picture.

\paragraph{Classical motivation.}
Let $\rho=\sum_x p(x)\proj{x}$ and let $\mathcal{E}(\sigma)=\sum_{x,y}P(y|x)\bra{x}\sigma\ket{x}\proj{y}$ be a classical channel. The operator
$\sum_{x,y}p(x)P(y|x)\,\proj{x}\otimes\proj{y}$ is the joint distribution $P(x,y)=P(y|x)p(x)$ written as a diagonal operator, and it equals the product $(\rho\otimes\mathds{1})J[\mathcal{E}]$, where $J[\mathcal{E}]=\sum_{x,y}P(y|x)\proj{x}\otimes\proj{y}$.
For quantum channels, $\rho\otimes\mathds{1}$ and $J[\mathcal{E}]$ do not commute in general, so this product is not Hermitian, and one symmetrizes.

\begin{definition}[Channel operator and state over time]\label{def:sot}
Let $\mathcal{E}:\mathcal{B}(\mathcal{H}_{S_0})\to\mathcal{B}(\mathcal{H}_{S_1})$ be linear. Its channel operator is
\begin{align}
    J[\mathcal{E}]=\sum_{i,j}\ket{i}\!\bra{j}\otimes\mathcal{E}(\ket{j}\!\bra{i})\in\mathcal{B}(\mathcal{H}_{S_0}\otimes\mathcal{H}_{S_1}),
\end{align}
i.e.\ the partial transpose on the input factor of the (unnormalized) Choi operator of Appendix~\ref{sec:pre}, with the factors ordered (input, output); the map $\mathcal{E}$ is recovered as $\mathcal{E}(\sigma)=\Tr_{S_0}[(\sigma\otimes\mathds{1})J[\mathcal{E}]]$.
For a state $\rho$ on $S_0$ and a channel $\mathcal{E}$, the \emph{state over time} is the Jordan product
\begin{align}
    \psi(\rho,\mathcal{E})=\mathrm{Jor}\big(\rho\otimes\mathds{1},\,J[\mathcal{E}]\big),\qquad \mathrm{Jor}(A,B)=\tfrac12 \{A,B\}.
\end{align}
\end{definition}

\begin{proposition}\label{prop:sot}
Let $\mathcal{E}$ be CPTP and $\rho$ a state. Then $\psi(\rho,\mathcal{E})$ is Hermitian with $\Tr\psi=1$, and
\begin{align}
    \Tr_{S_1}\psi(\rho,\mathcal{E})=\rho,\qquad \Tr_{S_0}\psi(\rho,\mathcal{E})=\mathcal{E}(\rho).
\end{align}
If $[\rho\otimes\mathds{1},J[\mathcal{E}]]=0$, then $\psi(\rho,\mathcal{E})=(\rho\otimes\mathds{1})J[\mathcal{E}]$.
\end{proposition}
\begin{proof}
Since $\mathcal{E}$ is CPTP, it is Hermiticity-preserving, so $J[\mathcal{E}]^\dagger=\sum_{ij}\ket{j}\!\bra{i}\otimes\mathcal{E}(\ket{j}\!\bra{i})^\dagger=\sum_{ij}\ket{j}\!\bra{i}\otimes\mathcal{E}(\ket{i}\!\bra{j})=J[\mathcal{E}]$, and the Jordan product of two Hermitian operators is Hermitian.
Trace preservation gives $\Tr_{S_1}J[\mathcal{E}]=\sum_{ij}\ket{i}\!\bra{j}\Tr(\ket{j}\!\bra{i})=\mathds{1}_{S_0}$, hence $\Tr_{S_1}[(\rho\otimes\mathds{1})J[\mathcal{E}]]=\Tr_{S_1}[J[\mathcal{E}](\rho\otimes\mathds{1})]=\rho$.
Moreover, $\Tr_{S_0}[(\rho\otimes\mathds{1})J[\mathcal{E}]]=\sum_{ij}\bra{j}\rho\ket{i}\,\mathcal{E}(\ket{j}\!\bra{i})=\mathcal{E}(\rho)$, and the same holds for the opposite order of the factors. The trace is then $\Tr\rho=1$.
\end{proof}

\noindent\textbf{Non-positivity.}
In general $\psi(\rho,\mathcal{E})\not\ge 0$, so it is a \emph{pseudo-density operator} (Definition~\ref{def:pdo} below) and not a state.
Its negative eigenvalues have been proposed as quantitative witnesses of quantum temporal correlations~\cite{Fitzsimons2015,Zhao2018} (see examples in Appendix~\ref{exm:tem}).
Ref.~\cite{Horsman2017} showed that no assignment defined on all pairs of operators can satisfy a set of natural requirements (Hermiticity, bilinearity, classical limit, marginals, associativity). The construction above avoids this no-go result because it is defined only on the data actually given, a state together with a channel~\cite{Fullwood2022}, and it is the unique solution of these requirements~\cite{LieNg2024,ParzygnatVirtual2024}.

\paragraph{Multi-time states over time.}
Let $S_0\xrightarrow{\mathcal{E}_1}S_1\xrightarrow{\mathcal{E}_2}\cdots\xrightarrow{\mathcal{E}_n}S_n$ be a chain of CPTP maps and $\rho_k=\mathcal{E}_k\circ\cdots\circ\mathcal{E}_1(\rho)$. Define recursively $\psi_0=\rho$ and
\begin{align}
    \psi_k=\mathrm{Jor}\big(\psi_{k-1}\otimes\mathds{1}_{S_k},\ \mathds{1}_{S_0\cdots S_{k-2}}\otimes J[\mathcal{E}_k]\big)\in\mathcal{B}(\mathcal{H}_{S_0}\otimes\cdots\otimes\mathcal{H}_{S_k}),\label{eq:sot_recursion}
\end{align}
and write $\psi(\rho,\mathcal{E}_1,\dots,\mathcal{E}_n)=\psi_n$.
The nontrivial result of Ref.~\cite{Fullwood2025} is that all $C_n=\frac{1}{n+1}\binom{2n}{n}$ ways of bracketing the iterated construction coincide, so $\psi_n$ is unambiguous (associativity). Equivalently,
$\psi_n=\mathrm{Jor}\big(\rho\otimes\mathds{1},J[\mathcal{E}_1]\otimes\mathds{1},\dots,\mathds{1}\otimes J[\mathcal{E}_n]\big)$ with the nested Jordan product $\mathrm{Jor}(A_0,\dots,A_n)=\mathrm{Jor}(A_0,\mathrm{Jor}(A_1,\dots))$.
For commuting (classical) data it reduces to $\sum_{\vec x}P(x_n|x_{n-1})\cdots P(x_1|x_0)p(x_0)\proj{\vec x}$.

\begin{proposition}[Marginals and quantum Chapman--Kolmogorov equation]\label{prop:sot_multi}
Let $\mathcal{E}_k:\mathcal{B}(\mathcal{H}_{S_{k-1}})\to\mathcal{B}(\mathcal{H}_{S_k})$, $k=1,\dots,n$, be CPTP maps, let $\rho$ be a state on $S_0$, and set $\rho_0:=\rho$ and $\rho_k:=\mathcal{E}_k(\rho_{k-1})$. Let $\psi_k$ be defined by the recursion~\eqref{eq:sot_recursion} with $\psi_0=\rho$ (for $k=1$ the factor $\mathds{1}_{S_0\cdots S_{k-2}}$ is absent). Then:
\begin{enumerate}[label=(\roman*)]
\item $\psi_n$ is Hermitian, $\Tr\psi_n=1$, and $\Tr_{S_{k+1}\cdots S_n}\psi_n=\psi_k$ for every $0\le k\le n-1$; in particular, $\Tr_{S_n}\psi_n=\psi_{n-1}$.
\item $\Tr_{\widehat{k}}\,\psi_n=\rho_k$ for every $0\le k\le n$, where $\Tr_{\widehat{k}}$ traces out all systems except $S_k$. Hence $\psi_n$ is a pseudo-density operator in the sense of Definition~\ref{def:pdo}.
\item For every $1\le k\le n$, the two-time marginal of $\psi_n$ on $S_{k-1}\otimes S_k$ equals $\psi(\rho_{k-1},\mathcal{E}_k)$.
\item $\Tr_{S_1}\psi(\rho,\mathcal{E}_1,\mathcal{E}_2)=\psi(\rho,\mathcal{E}_2\circ\mathcal{E}_1)$.
\end{enumerate}
\end{proposition}

\begin{proof}
Write $J_k:=J[\mathcal{E}_k]$, which acts on $S_{k-1}\otimes S_k$. The proof uses only three elementary facts.
(a) If an operator $A$ acts as the identity on a subsystem $K$, then $\Tr_K[AX]=A\,\Tr_K X$ and $\Tr_K[XA]=(\Tr_K X)\,A$ for every operator $X$.
(b) Trace preservation gives $\Tr_{S_k}J_k=\sum_{i,j}\ket{i}\!\bra{j}\,\Tr\,\mathcal{E}_k(\ket{j}\!\bra{i})=\mathds{1}_{S_{k-1}}$.
(c) Since $\mathcal{E}_k$ is Hermiticity-preserving, $J_k^\dagger=J_k$ (see the proof of Proposition~\ref{prop:sot}), and the Jordan product of two Hermitian operators is Hermitian.

\emph{(i)} Hermiticity of $\psi_k$ follows from (c) by induction on $k$, starting from $\psi_0=\rho$. The operator $\psi_{k-1}\otimes\mathds{1}_{S_k}$ acts as the identity on $S_k$, so (a) and (b) give
\begin{align}
&\Tr_{S_k}\!\big[(\psi_{k-1}\otimes\mathds{1}_{S_k})(\mathds{1}_{S_0\cdots S_{k-2}}\otimes J_k)\big]\nonumber\\
&\qquad=\psi_{k-1}\big(\mathds{1}_{S_0\cdots S_{k-2}}\otimes\Tr_{S_k}J_k\big)=\psi_{k-1},
\end{align}
and the same holds for the product in the opposite order. Hence $\Tr_{S_k}\psi_k=\psi_{k-1}$. Iterating from $S_n$ down to $S_{k+1}$ gives $\Tr_{S_{k+1}\cdots S_n}\psi_n=\psi_k$, and taking the full trace gives $\Tr\psi_n=\Tr\psi_0=\Tr\rho=1$.

\emph{(ii) and (iii)} By (i), the marginals of $\psi_n$ on $S_k$ and on $S_{k-1}\otimes S_k$ coincide with the corresponding marginals of $\psi_k$. It therefore suffices to show that $\Tr_{S_0\cdots S_{k-2}}\psi_k=\psi(\rho_{k-1},\mathcal{E}_k)$ and $\Tr_{\widehat{k}}\,\psi_k=\rho_k$ for $1\le k\le n$; the case $k=0$ of (ii) is simply $\psi_0=\rho_0$. We proceed by induction on $k$ and assume $\Tr_{\widehat{k-1}}\,\psi_{k-1}=\rho_{k-1}$, which holds for $k=1$. The second operand in Eq.~\eqref{eq:sot_recursion} acts as the identity on $S_0\cdots S_{k-2}$, so by (a)
\begin{align}
\Tr_{S_0\cdots S_{k-2}}\psi_k
&=\mathrm{Jor}\big((\Tr_{S_0\cdots S_{k-2}}\psi_{k-1})\otimes\mathds{1}_{S_k},\,J_k\big)\nonumber\\
&=\mathrm{Jor}\big(\rho_{k-1}\otimes\mathds{1}_{S_k},\,J_k\big)=\psi(\rho_{k-1},\mathcal{E}_k),
\end{align}
where we used $\Tr_{S_0\cdots S_{k-2}}\psi_{k-1}=\Tr_{\widehat{k-1}}\,\psi_{k-1}=\rho_{k-1}$. This proves (iii). Tracing out $S_{k-1}$ and applying Proposition~\ref{prop:sot} gives $\Tr_{\widehat{k}}\,\psi_k=\mathcal{E}_k(\rho_{k-1})=\rho_k$, which closes the induction and proves (ii). Since $\psi_n$ is Hermitian and each single-time marginal $\rho_k$ is a state, $\psi_n$ is a pseudo-density operator.

\emph{(iv)} Put $a=\rho\otimes\mathds{1}_{S_1}\otimes\mathds{1}_{S_2}$, $b=J_1\otimes\mathds{1}_{S_2}$, and $c=\mathds{1}_{S_0}\otimes J_2$, so that $\psi_2=\mathrm{Jor}(\mathrm{Jor}(a,b),c)=\tfrac14\{\{a,b\},c\}$. Since $a$ acts on $S_0$ and $c$ on $S_1\otimes S_2$, we have $[a,c]=0$, and therefore
\begin{align}
\{\{a,b\},c\}&=abc+bac+cab+cba\nonumber\\
&=abc+bca+acb+cba=\{a,\{b,c\}\},
\end{align}
where we used $bac=bca$ and $cab=acb$. Write $\mathcal{E}_1(\ket{j}\!\bra{i})=\sum_{l,k}X^{ji}_{lk}\ket{l}\!\bra{k}$ and $J_2=\sum_{p,q}\ket{p}\!\bra{q}\otimes\mathcal{E}_2(\ket{q}\!\bra{p})$. Multiplying out,
\begin{align}
bc=\sum_{i,j,k,l,q}X^{ji}_{lk}\,\ket{i}\!\bra{j}\otimes\ket{l}\!\bra{q}\otimes\mathcal{E}_2(\ket{q}\!\bra{k}),
\end{align}
and the trace over $S_1$ sets $q=l$:
\begin{align}
\Tr_{S_1}[bc]&=\sum_{i,j}\ket{i}\!\bra{j}\otimes\mathcal{E}_2\Big(\sum_{l,k}X^{ji}_{lk}\ket{l}\!\bra{k}\Big)\nonumber\\
&=\sum_{i,j}\ket{i}\!\bra{j}\otimes(\mathcal{E}_2\circ\mathcal{E}_1)(\ket{j}\!\bra{i})=J[\mathcal{E}_2\circ\mathcal{E}_1].
\end{align}
The same computation gives $\Tr_{S_1}[cb]=J[\mathcal{E}_2\circ\mathcal{E}_1]$, so $\Tr_{S_1}\{b,c\}=2J[\mathcal{E}_2\circ\mathcal{E}_1]$. Finally, $a$ acts as the identity on $S_1$, so by (a)
\begin{align}
\Tr_{S_1}\psi_2&=\tfrac14\Big[(\rho\otimes\mathds{1}_{S_2})\,\Tr_{S_1}\{b,c\}\nonumber\\
&\qquad\ +\Tr_{S_1}\{b,c\}\,(\rho\otimes\mathds{1}_{S_2})\Big]\nonumber\\
&=\mathrm{Jor}\big(\rho\otimes\mathds{1}_{S_2},\,J[\mathcal{E}_2\circ\mathcal{E}_1]\big)\nonumber\\
&=\psi(\rho,\mathcal{E}_2\circ\mathcal{E}_1).
\end{align}
\end{proof}
Proposition~\ref{prop:sot_multi} shows that the state over time is the quantum counterpart of the joint distribution of a classical Markov chain. Item (ii) states that every single-time marginal of $\psi_n$ is the actual state of the system at that time, so $\psi_n$ is a consistent pseudo-density operator even though it is, in general, not positive. Item (iii) states that each two-time marginal depends only on the state at the earlier time and on the one-step channel, so $\psi_n$ is assembled entirely from one-step data. Item (iv) is the operator analogue of the Chapman--Kolmogorov equation~\eqref{eq:CK_eq}: tracing out an intermediate time composes the channels, just as summing over $x_2$ composes the transition probabilities. There is, however, an important difference. In the classical case, Eq.~\eqref{eq:CK_eq} is derived from the Markov condition~\eqref{classical_markov}, whereas (iv) holds identically for every chain of channels, because $\psi_n$ contains no information beyond $(\rho,\mathcal{E}_1,\dots,\mathcal{E}_n)$. The state over time is therefore Markovian by construction, and the quantum Chapman--Kolmogorov equation is a consistency property of $\psi_n$ rather than a test of Markovianity. Accordingly, the negative eigenvalues of $\psi_n$ witness quantum temporal correlations of the system itself, not environmental memory (cf.\ Example~\ref{ex:sot_id}). Finally, the proof uses only the trace preservation and Hermiticity preservation of the maps $\mathcal{E}_k$, and never their complete positivity. Items (i)--(iv) therefore remain valid for chains of Hermiticity-preserving and trace-preserving propagators, such as the intermediate maps $V_{t_k,t_{k-1}}=\Lambda_{t_k}\circ\Lambda_{t_{k-1}}^{-1}$ of an invertible but CP-indivisible dynamics (Sec.~\ref{subsec:RHP_QNM}), for which the single-time marginals $\rho_k=\Lambda_{t_k}(\rho)$ are still states. The correspondence between pseudo-density operators and processes discussed below, however, relies on complete positivity and does not extend to this case.\\

\paragraph{Pseudo-density operators and extraction of the dynamics.}

\begin{definition}[Pseudo-density operator]\label{def:pdo}
A Hermitian $\tau\in\mathcal{B}(\mathcal{H}_{S_0}\otimes\cdots\otimes\mathcal{H}_{S_n})$ is a pseudo-density operator if every single-time marginal $\Tr_{\widehat{k}}\tau$, $k=0,\dots,n$, is a state. In particular $\Tr\tau=1$, but $\tau\ge0$ is not required.
\end{definition}

By Proposition~\ref{prop:sot_multi}(ii), every state over time is a pseudo-density operator. Conversely, Ref.~\cite{Fullwood2025} shows that, whenever the input marginal $\rho=\Tr_{S_1}\tau=\sum_a r_a\proj{a}$ is invertible, the dynamics can be read off from $\tau$ uniquely: the equation $\tfrac12\{\rho\otimes\mathds{1},X\}=\tau$ has a unique solution~\cite{Fullwood2025}, which in the eigenbasis of $\rho$ reads
\begin{align}
X_\tau=\sum_{a,b}\ket{a}\!\bra{b}\otimes\frac{2\,\tau_{ab}}{r_a+r_b},\qquad \tau_{ab}=(\bra{a}\otimes\mathds{1})\,\tau\,(\ket{b}\otimes\mathds{1}),
    \label{eq:sot_inverse}
\end{align}
with candidate dynamics $\mathcal{E}_\tau(\sigma)=\Tr_{S_0}[(\sigma\otimes\mathds{1})X_\tau]$. Hermiticity and trace preservation of $\mathcal{E}_\tau$ hold automatically; only complete positivity can fail, and it holds if and only if $X_\tau^{T_{S_0}}\ge0$. Thus, $(\rho,\mathcal{E})\mapsto\psi(\rho,\mathcal{E})$ is a bijection between pairs of an invertible state and a channel, and the pseudo-density operators with invertible input marginal whose $\mathcal{E}_\tau$ is completely positive. Invertibility is essential: a state over time carries no information about how the channel acts on inputs that $\rho$ never populates. For $n$ steps, if every $\rho_{k}$ is invertible, a state over time $\psi_n$ is fixed by its nearest-neighbour two-time marginals: by Proposition~\ref{prop:sot_multi}(iii) each marginal equals $\psi(\rho_{k-1},\mathcal{E}_k)$, from which $\mathcal{E}_k$ is recovered by Eq.~\eqref{eq:sot_inverse}. This mirrors how the joint distribution of a classical Markov chain is fixed by its initial distribution and transition probabilities (see the examples in Appendix~\ref{exm:tem}).

The state over time depends on the dynamics only through the data $(\rho,\mathcal{E}_1,\dots,\mathcal{E}_n)$, and, by the correspondence established above, it determines these data uniquely. It therefore encodes exactly the information contained in the reduced one-step maps and nothing beyond it. In this way, it provides a faithful, basis-independent description of the temporal correlations that the system itself carries between times, in direct analogy with the joint distribution of a classical Markov chain. At the same time, as with the two-time maps of                                                                                      Sec.~\ref{subsec:RHP_QNM}, the way these maps are generated lies outside its scope: any two processes that share the same reduced propagators are assigned the same $\psi$, irrespective of whether the underlying process is Markovian or retains memory in the environment. The state over time is therefore not designed to distinguish a process with memory from a Markovian one with the same reduced maps, nor classical from quantum memory. Such distinctions concern the multi-time response of the process to interventions, which is captured by the process tensor rather than by $\psi$.

For instance, in the model of Appendix~\ref{exam:exmp_pro_ten}, the process tensor satisfies $T_{2:0}\neq C_{t_2:t_1}\otimes C_{t_1:t_0}$. Provided the propagator $\Lambda_{t_2:t_1}$ is CPTP (which holds there for $\cos^2\theta\ge1/5$), the Markovian process with the factorized form has the same reduced maps $\Lambda_{t_1:t_0},\Lambda_{t_2:t_1}$, and hence the same $\psi$. As Example~\ref{ex:sot_id} shows, the negative eigenvalues of $\psi$ quantify the non-classicality of the system's own temporal correlations, not environmental memory. Relating pseudo-density operators to general process tensors with memory, and their negativity to the temporal entanglement of Sec.~\ref{subsec:GC_quantum_memory}, requires information beyond the reduced maps and is not pursued here.

\section{Process tensors vs.\ dynamical maps}\label{sec:PT_vs_DM}
A sharp and practically important question in the study of genuine quantum
non-Markovianity is whether the memory sustaining non-Markovian dynamics is
itself quantum, or whether a classical stochastic process suffices to reproduce
it. B\"{a}cker, Link, and Strunz~\cite{backer_2025,Link_2024} address this directly for
the spin-boson model by deploying two local quantum memory criteria that differ
in the amount of dynamical information they require. Process tensors and
dynamical maps are computed via a numerically exact matrix-product-operator
(MPO) representation, giving access
to the full multi-time propagator in the strongly coupled regime across a broad
range of bath parameters and temperatures.

\subsection{Classical vs.\ quantum memory}

A process possesses \emph{classical memory} if its environment can be simulated by a stochastic classical variable. Operationally, this means the single-intervention process tensor $T_{t_1,t_2}$ must admit a separable Choi-state decomposition across the cut separating the first step (from $t_0$ to $t_1$) from the second step (from $t_1$ to $t_2$),
\begin{align}
    \tau_{T} = \sum_k p_k\, \rho_k^{\mathrm{in}} \otimes \sigma_k^{\mathrm{out}},
    \qquad p_k \geq 0, \quad \sum_k p_k = 1. \label{eq:classical_memory}
\end{align}
Entanglement of $\tau_{T}$ across this bipartition therefore certifies that classical memory is insufficient: the
environment must maintain genuine quantum coherences between $t_1$ and $t_2$. Processes with quantum memory form a strict subset of all non-Markovian processes (classical-memory processes, which include the Markovian ones, form the complement within all processes), so detecting quantum memory is a strictly stronger statement than detecting
non-Markovianity alone.

\subsection{Two criteria for quantum memory}

\paragraph{Criterion 1: Process-tensor (PT) criterion.} With access to the single-intervention process tensor $T_{t_1,t_2}$,
quantum memory is witnessed by a positive lower bound on the I-concurrence of
its Choi state $\tau_{T}$,
\begin{align}
    \mathcal{C}_{<}(\tau_{T}) > 0
    \quad \Longrightarrow \quad
    \text{quantum memory.}
\label{eq:PT_criterion}
\end{align}
This quantity is computed from the singular values of the partially transposed Choi matrix and is directly accessible from the MPO-generated process tensor at any pair of intervention times $(t_1, t_2)$.

\paragraph{Criterion 2: Dynamical-map (DM) criterion.}
Given only the pair of CPTP maps $\mathcal{E}_{t_1}$ and $\mathcal{E}_{t_2}$, i.e., the time marginals of the process tensor, the classical-memory hypothesis constrains the Choi states
   $\chi_{t_i}=(\mathcal E_{t_i}\otimes\mathds 1_{\rm anc})[|\Psi^+\rangle\langle\Psi^+|]$, $i=1,2$,
   where $|\Psi^+\rangle$ is a maximally entangled system--ancilla state. Quantum memory is
   certified whenever
   \begin{align}
   \mathcal{C}_{<}(\chi_{t_2}) > \sqrt{2\big(1-\mathrm{Tr}[(\mathrm{Tr}_{\rm anc}\chi_{t_1})^2]\big)} ,
   \label{eq:DM_criterion}
   \end{align}
which uses the dynamical maps alone. Because $\mathcal{E}_{t_1}$ and $\mathcal{E}_{t_2}$ are marginals of $T_{t_1,t_2}$, the DM criterion discards the temporal correlations retained in the full process tensor, so it can never use more information than the PT criterion; in the cases studied it is found to be considerably weaker~\cite{backer_2025}.
\subsection{Comparison}
The central finding is a clear operational hierarchy between the two criteria. For resonant (Lorentzian) bath environments at low temperatures, both criteria detect quantum memory, but the PT criterion does so across a far wider parameter
window; the DM criterion requires fine-tuned conditions of long memory times and near-resonance to succeed. For dissipative (Ohmic) baths, the dynamics is non-Markovian even with respect to the more restrictive criterion of $P$-indivisibility, yet the DM criterion fails to certify quantum memory in any explored parameter regime, even at zero temperature, while the PT criterion still succeeds. The failure of the DM criterion is therefore not a consequence of weak non-Markovianity, but a separate phenomenon~\cite{backer_2025}.
This demonstrates that neither CP- nor $P$-indivisibility is sufficient for quantum memory to be detectable at the dynamical-map level. The Choi-state entanglement of the process tensor is therefore the more powerful operational quantity.

Here, ``DM-detectable'' (``PT-detectable'') quantum memory denotes quantum memory that
can in principle be certified from the dynamical maps alone (from the single-intervention
process tensor). Since the maps are marginals of the process tensor, the former class is
contained in the latter, and the specific criteria~\eqref{eq:DM_criterion}
and~\eqref{eq:PT_criterion} are sufficient tests within each class. The results establish
the following inclusions among the corresponding classes of processes:
\begin{align}
   & \text{DM-detectable quantum memory}
    \;\subset\;
    \text{PT-detectable quantum memory}
    \nonumber \\ &\hspace{0.8cm}\;\subseteq\;
    \text{quantum memory}
    \;\subset\;
    \text{non-Markovian},
    \label{eq:hierarchy}
\end{align}
together with $\text{$P$-indivisible}\subset\text{CP-indivisible}$. There is no general inclusion between the divisibility-based classes and the quantum-memory classes: in the Ohmic case, the dynamics is $P$-indivisible, yet its quantum memory is detected by the PT criterion but not by the DM criterion. Detecting more quantum memory requires progressively
more multi-time information: the dynamical map encodes only single-time
statistics, while the process tensor exposes temporal correlations between interventions at $t_1$ and $t_2$. The practical upshot is that
single-intervention process tomography is a considerably more reliable tool for certifying genuine quantum non-Markovianity in such systems, whereas
dynamical-map tomography alone can miss it entirely, particularly for
dissipative environments in the spin-boson model at low temperature~\cite{backer_2025}. This work thus provides a concrete
operational protocol for distinguishing genuine quantum memory from classical non-Markovian effects, directly relevant to quantum information processing,
error correction, and the broader program of characterizing genuine quantum
non-Markovianity reviewed in this article.

\section{Summary}
In this review, we presented a comprehensive and unified account of non-Markovian dynamical systems, with a particular focus on the nature of the backflow behavior. While non-Markovianity is often associated with information backflow or the violation of divisibility conditions, an emerging question arising from recent research is whether such signatures alone determine whether the cause of non-Markovianity genuinely arises from a quantum origin or can have a classical explanation in information theory. The main objective of this article has been to organize and compare the different approaches developed in this direction for characterizing non-Markovianity and clarify which aspects of backflow effects lead to genuine quantum non-Markovianity.

We began by restating the notion of classical non-Markovianity, where memory effects arise when the future evolution of a stochastic process depends on its past history beyond the most recent state. In classical dynamics, such behavior is identified through violations of the Markov condition (a violation of the Chapman--Kolmogorov equation is sufficient, but not necessary, to detect it), and the memory originates from classical correlations or hidden variables that retain information about earlier states.
This classical perspective provides a natural baseline for understanding quantum non-Markovian dynamics, where similar temporal correlations may appear but do not necessarily imply a quantum origin. Building on this foundation, we then reviewed the principal frameworks used to characterize quantum non-Markovianity in open quantum systems. Structural approaches based on divisibility and operational approaches based on trace distance and the revival of correlations, etc., identify Markovian dynamics and non-Markovian behavior. Information-theoretic formulations further connect these ideas to mutual and conditional mutual information, linking non-Markovianity to violations of data-processing inequalities. A more general contractive framework unifies these perspectives by showing that any increase in a contractive quantity during the evolution provides a signature of memory effects in the dynamics. We have also reviewed the fundamental limitations of two-time descriptions.
Memory is inherently a multi-time phenomenon, and the full structure of temporal correlations can only be captured through the process-tensor formalism. In this framework, Markovian processes are those whose process tensor factorizes into a product of independent one-step channels. Any deviation from this factorized structure reveals multi-time correlations between different stages of the evolution and therefore indicates non-Markovian dynamics. Importantly, this framework shows that CP-divisible dynamics can still possess temporal correlations and thus fail to be Markovian at the level of the full process.

Despite the many methods available to detect non-Markovianity, most of them do not reveal the origin of the backflow of information. In particular, information backflow or correlation revivals may occur even when the environment stores only classical information. An example  discussed in this review (in Appendix~\ref{exam:mixing}) is the convex mixing of Markovian dynamical maps, where classical uncertainty about the applied dynamics leads to apparent information backflow. In such cases, the memory is stored in classical degrees of freedom and does not involve quantum coherence or entanglement. This observation highlights the need for refined criteria capable of distinguishing classical memory effects from genuinely quantum ones. We have discussed recently developed approaches to addressing this problem. Distinguishability-based methods introduce quantum memory witnesses that rule out classical hidden-variable explanations of information backflow. Criteria based on the entanglement structure of Choi states provide a way to detect quantum memory directly from dynamical maps. Operational multi-time approaches characterize the nature of memory through the classical simulability of sequential measurement statistics. In the process-tensor framework, the distinction becomes structural: classical memory implies a separable process tensor that can be decomposed into mixtures of independent temporal channels, whereas temporal entanglement of the process tensor certifies quantum memory. Another important development reviewed in this work is the conceptual distinction between information revivals and genuine information backflow. By identifying and excluding such noncausal information revivals, one isolates the genuine information backflow. Applied to the quantum switch, this criterion shows that, in the examples studied, the revivals induced by the standard switch are non-causal and hence do not certify genuine quantum non-Markovianity, whereas coherently ordering interactions with a shared environment can produce revivals that resist every inert extension tested

We also reviewed approaches that quantify temporal correlations directly: temporal steering provides an ancilla-free witness of strong non-Markovianity, and states over time (pseudo-density operators) give a faithful single-operator description of the temporal correlations carried by the system, which, however, is insensitive to environmental memory by construction.

We have also discussed the comparison between the process tensor and the dynamical map introduced by Bäcker \emph{et al.}~\cite{backer_2025}, who certified genuine quantum memory with both approaches and showed that the process-tensor
criterion strictly outperforms the dynamical-map criterion, establishing
that CP-indivisibility alone is insufficient to distinguish quantum from
classical environmental memory, and that single-intervention process
tomography is a more powerful diagnostic for genuine quantum non-Markovianity.

Finally, the review discussed a perspective in which non-Markovianity is analyzed at the level of quantum states rather than dynamics. The concept of squashed quantum non-Markovianity provides a measure of genuine quantum non-Markovianity by removing contributions that can be explained through classical mixture. This framework establishes a clear separation between non-genuine and genuinely quantum non-Markovian correlations and shows that genuine quantum non-Markovianity behaves as a well-defined quantum resource, satisfying properties such as convexity, monogamy, and additivity.

Taken together, the developments reviewed here show that quantum non-Markovianity is a layered concept. Detecting memory effects, observing information revivals, and identifying genuine backflow of information correspond to different levels of analysis that require distinct theoretical tools. By organizing these approaches within a common information-theoretic framework, this review clarified the conceptual structure of non-Markovian dynamics and highlighted the emerging understanding that only backflow of information that cannot be simulated by classical explanations should be regarded as genuine backflow, leading to genuine quantum non-Markovian dynamics. These insights
potentially
provide a clearer foundation for future research on non-Markovian processes and their applications in quantum information science and quantum technologies.

\section*{Declaration of generative AI}
During the preparation of this work, the author(s) used Claude (Anthropic) most of the time to improve the language and sometimes to find suitable literature. The author(s) reviewed and edited the output as needed and take full responsibility for the content of the published article.

\begin{acknowledgments}
R.G. acknowledges the funding support from the Israel Science
Foundation under Grant No. 1192/24. U.S.
acknowledges financial support from the Anusandhan National Research Foundation (ANRF), Government of India,
under the Grant No. ANRF/ARG/2025/004617/PS.
\end{acknowledgments}

\bibliography{library}

\appendix
\renewcommand{\thesubsection}{\thesection.\arabic{subsection}}
\numberwithin{equation}{section}
\numberwithin{figure}{section}

\section{Preliminaries}\label{sec:pre}
In this section, we briefly introduce the mathematical tools from quantum information theory that will be used throughout the review. These notions provide the formal framework for describing open quantum dynamics, correlations, and information-theoretic quantities that characterize non-Markovian processes.

A quantum system associated with a Hilbert space $\mathcal{H}$ is described by a density operator $\rho\in\mathcal{B}(\mathcal{H})$, where $\mathcal{B}(\mathcal{H})$ denotes the space of bounded linear operators acting on $\mathcal{H}$. A valid quantum state is a positive semi-definite operator with unit trace ($\rho \ge 0,\ \  \Tr(\rho)=1$). Pure states correspond to projectors of the form $\rho=\ket{\psi}\!\bra{\psi}$, while mixed states arise as convex combinations of pure states. For composite systems described on a tensor-product Hilbert space $\mathcal{H}_A\otimes\mathcal{H}_B$, the reduced state of a subsystem is obtained through the partial trace operation, for example
\begin{align}
\rho_A = \Tr_B(\rho_{AB}).
\end{align}
The most general physical evolution of an open quantum system is described by a completely positive and trace-preserving (CPTP) linear map $\Lambda:\mathcal{B}(\mathcal{H}_S)\rightarrow \mathcal{B}(\mathcal{H}_S),$ commonly referred to as a quantum channel. Complete positivity ensures that the map remains positive even when extended with an arbitrary ancilla system. By the Kraus representation theorem, any CPTP map can be written in the operator sum form
\begin{align}
\Lambda(\rho)=\sum_i K_i \rho K_i^\dagger ,
\end{align}
where the operators $\{K_i\}$ satisfy the normalization condition $\sum_i K_i^\dagger K_i = \mathds{1}$. This representation provides a convenient description of the reduced dynamics of a system interacting with an environment.

An equivalent characterization of quantum channels is provided by the Choi--Jamio{\l}kowski isomorphism, which establishes a one-to-one correspondence between linear maps and bipartite operators. For a channel $\Lambda$ acting on a $d$-dimensional system, the associated Choi operator is defined as
\begin{align}
C[\Lambda] = (\Lambda\otimes\mathds{1}) \left( \ket{\phi^+}\ \bra{\phi^+}\right),
\end{align}
where $\ket{\phi^+}=\sum_{i=0}^{d-1} \ket{i}\ket{i}$ is an unnormalized maximally entangled state on $\mathcal{H}_S\otimes\mathcal{H}_A$ (in the main text, the normalized version $\ket{\phi^+}/\sqrt{d}$ is used whenever the Choi \emph{state} is meant). The channel $\Lambda$ is completely positive if and only if the corresponding Choi operator is positive semi-definite. This representation plays a central role in several constructions discussed later, particularly in the characterization of quantum memory through dynamical maps.

Information-theoretic properties of quantum states are quantified using entropy measures. The von Neumann entropy of a state $\rho$ is defined as
\begin{align}
S(\rho)=-\Tr(\rho\log\rho).
\end{align}
For a bipartite state $\rho_{AB}$, the total correlations shared between subsystems $A$ and $B$ are quantified by the quantum mutual information
\begin{align}
I(A:B)_\rho = S(\rho_A)+S(\rho_B)-S(\rho_{AB}).
\end{align}
More generally, for a tripartite state $\rho_{ABC}$ one defines the quantum conditional mutual information (QCMI)
\begin{align}
I(A;C|B)_\rho = S(AB)_\rho + S(BC)_\rho - S(B)_\rho - S(ABC)_\rho .
\end{align}
The QCMI plays a central role in quantum information theory because it characterizes the structure of quantum Markov states and satisfies the strong subadditivity inequality $I(A;C|B)_\rho \ge 0 $. Several notions of quantum non-Markovianity discussed in this review can be naturally formulated in terms of the behavior of these correlation measures under dynamical evolution.

Another important quantity used to quantify distinguishability between quantum states is the trace distance. For an operator $X$, the trace norm is defined as
\begin{align}
\|X\|_1=\Tr\sqrt{X^\dagger X}.
\end{align}
The trace distance between two states $\rho$ and $\sigma$ is then
\begin{align}
D(\rho,\sigma)=\frac12\|\rho-\sigma\|_1 ,
\end{align}
which has a direct operational interpretation in state discrimination: the optimal probability of correctly distinguishing $\rho$ and $\sigma$, given with equal prior probabilities, is $P_{\rm succ}=\tfrac12\left[1+D(\rho,\sigma)\right]$.

More generally, distinguishability and correlation measures often satisfy a contractivity property under physical quantum evolutions. A function $f(\rho,\sigma)$ defined on pairs of states is called contractive if for every CPTP map $\Phi$,
\begin{align}
f\bigl(\Phi(\rho),\Phi(\sigma)\bigr)\le f(\rho,\sigma).
\end{align}
This property expresses the intuitive principle that information cannot increase under local physical operations acting on the system. Contractive quantities play an important role in several operational definitions of non-Markovianity, where temporary violations of monotonic decay are interpreted as signatures of information flowing back to the system from its environment.

\section{Examples}\label{Example_1}
\subsection{Example: RHP vs BLP non-Markovianity}\label{exmpl_RHP-BLP}
Here, we show that the BLP and RHP approaches capture different properties and are not equivalent. For this example, it is convenient to start from the simplest class of qubit channels, the Pauli channels. A Pauli channel is a probabilistic mixture of Pauli conjugations (not every qubit channel is of this form; e.g., amplitude damping is not),
\begin{align}
\Lambda_t(\rho)=\sum_{i=0}^{3}p_i(t)\,\sigma_i\rho\sigma_i,
\end{align}
where the coefficients $p_i(t)$ are ordinary probabilities satisfying $p_i(t)\ge 0$ and $\sum_{i}p_i(t)=1$. If a qubit state is written in Bloch form,
\begin{align}
\rho=\tfrac12(I+r_1\sigma_x+r_2\sigma_y+r_3\sigma_z),
\end{align}
the action of $\Lambda_t$ on the Bloch vector becomes diagonal, $r_i(t)=\eta_i(t)\,r_i(0)$, where each $\eta_i(t)$ is an explicit linear combination of the probabilities $p_i(t)$. Differentiating this evolution gives
\begin{align}
\dot{r}_i(t)=\frac{\dot{\eta}_i(t)}{\eta_i(t)}r_i(t)\equiv \lambda_i(t)\, r_i(t),
\end{align}
which expresses the instantaneous rates of contraction of the Bloch components. It is a standard result that any time-local generator with such diagonal Bloch dynamics can be rewritten in the canonical (diagonal) Lindblad form
\begin{align}
    \dot{\rho}(t)=\sum_{i=1}^3\gamma_i(t)\bigl(\sigma_i\rho(t)\sigma_i-\rho(t)\bigr),
\end{align}
where the combinations $(\sigma_i\rho\sigma_i-\rho)$ ensure trace preservation and the three functions $\gamma_i(t)$ are uniquely determined by the three Bloch contraction rates $\lambda_i(t)$. Thus, the coefficients $\gamma_i(t)$ are not chosen arbitrarily; they arise naturally when one rewrites a general Pauli-channel qubit evolution in the Lindblad representation.

A particularly illuminating choice of the decay rates, originally introduced by Hall \emph{et al}~\cite{Hall2014}. (there with an overall factor $1/2$ in the generator), is $\gamma_1(t)=\gamma_2(t)=1$ and $\gamma_3(t)=-\tanh t$. Because $\tanh t>0$ for all $t>0$, the third rate is negative at every instant, meaning that the intermediate maps $\Lambda_{t+dt,t}$ are never completely positive. Therefore, the RHP criterion based on CP-divisibility identifies the dynamics as non-Markovian. To visualize the actual evolution, we write the state in Bloch form,
\begin{align}
\rho(t)=\frac12\!\left(I+r_1(t)\sigma_x+r_2(t)\sigma_y+r_3(t)\sigma_z\right),
\end{align}
and use the identity $\sigma_i\,\rho(t)\,\sigma_i-\rho(t)
    =-\sum_{j\neq i} r_j(t)\,\sigma_j$,
which leads directly to the Bloch equations
\begin{align}
&\dot{r}_1(t)=-2\!\left[\,1-\tanh(t)\,\right] r_{1}(t),\nonumber\\
&\dot{r}_2(t)=-2\!\left[\,1-\tanh(t)\,\right]r_{2}(t),\nonumber\\
&\dot{r}_3(t)=-4\,r_3(t).
\end{align}
Solving these differential equations (using $\int_0^t\tanh s\,ds=\ln\cosh t$) yields
\begin{align}
r_i(t)=
\begin{cases}
e^{-2t}\cosh^2\! t\ \,r_i(0), & i=1,2,\\[6pt]
e^{-4t}\,r_3(0), & i=3.
\end{cases}
\end{align}
So the dynamical map contracts each Bloch component with the strictly decreasing factors $\eta_1(t)=\eta_2(t)=e^{-2t}\cosh^2 t$ and $\eta_3(t)=e^{-4t}$. The trace distance between any two states with Bloch vectors $\mathbf{r}$ and $\mathbf{s}$ evolves as
\begin{align}
D[\mathbf{r},\mathbf{s}](t)=\tfrac12\sqrt{\eta_1(t)^2\Delta r_1^2+\eta_2(t)^2\Delta r_2^2+\eta_3(t)^2\Delta r_3^2},
\end{align}
with $\Delta r_i=r_i-s_i$, and since all $\eta_i(t)$ decrease monotonically, the trace distance is strictly monotonically decreasing as well (for $\mathbf{r}\neq\mathbf{s}$).

\subsection{Example: process tensor non-Markovianity} \label{exam:exmp_pro_ten}
Consider a dynamical model in which a system qubit $S$ interacts sequentially with two environment qubits $E_1$ and $E_2$ at times $t_1$ and $t_2$. Each dynamics is governed by the same unitary $U$ acting on $S\otimes E$; a convenient choice is the partial-swap unitary
\begin{align}
U(\theta) = \cos\theta \, \mathds{1}
        + i\sin\theta \, U_\mathrm{SWAP}\;=\;e^{i\theta U_{\rm SWAP}},
\end{align}
which generates a CPTP map on $S$ when $E$ is traced out. Let the joint initial state of the environment be
\begin{align}
\rho_{E_1E_2}
= \frac{1}{2}\bigl(|00\rangle\!\langle 00| + |11\rangle\!\langle 11|\bigr),
\label{eq:CorrEnv}
\end{align}
which contains classical correlations between $E_1$ and $E_2$. The system has an arbitrary initial state $\rho_S(t_0)$. At time $t_1$, $S$ interacts with $E_1$ via $U(\theta)$ while $E_2$ is untouched:
\begin{align}
\rho_{SE_1E_2}(t_1)= (U_{SE_1}\otimes \mathds{1}_{E_2})
\bigl[\rho_S(t_0)\otimes\rho_{E_1E_2}\bigr]
       (U_{SE_1}^\dagger\otimes \mathds{1}_{E_2}).
\end{align}
Tracing out $E_1$ yields the reduced state of the system:
\begin{align}
\rho_S(t_1)
    = \Lambda_{t_1:t_0}[\rho_S(t_0)],
\end{align}
where $\Lambda_{t_1:t_0}$ is a CPTP map. Explicitly, with $E_1$ in $\ket{k}$ the reduced map is $\Lambda^{(k)}(\rho)=\cos^2\theta\,\rho+\sin^2\theta\,\proj{k}+i\sin\theta\cos\theta\,[\proj{k},\rho]$, and averaging over $k$ gives the depolarizing map $\Lambda_{t_1:t_0}(\rho)=\cos^2\theta\,\rho+\sin^2\theta\,\mathds{1}/2$, which is invertible for $\cos\theta\neq0$. At time $t_2$, the system interacts with the second environment system $E_2$:
\begin{align}
\rho_{SE_2}(t_2)
    &= U_{SE_2}
       \bigl[\,\rho_{SE_2}(t_1)\,\bigr]
       U_{SE_2}^\dagger .
\end{align}
Crucially, because $E_1$ and $E_2$ were initially correlated as in~\eqref{eq:CorrEnv}, the reduced joint state $\rho_{SE_2}(t_1)=\tfrac12\sum_k\Lambda^{(k)}[\rho_S(t_0)]\otimes\proj{k}_{E_2}$ before the second dynamics is not of the product form $\rho_S(t_1)\otimes\rho_{E_2}$, and it explicitly depends on the earlier interaction with $E_1$. Thus, the second dynamics is influenced by the history, even though it involves a different environmental particle. Since $\Lambda_{t_1:t_0}$ is invertible, the intermediate map $\Lambda_{t_2:t_1}=\Lambda_{t_2:t_0}\circ\Lambda_{t_1:t_0}^{-1}$ is well defined. A direct calculation shows that it is the Pauli channel with Bloch contraction factors $\eta_1=\eta_2=\cos2\theta$ and $\eta_3=\cos^2\theta$, which is completely positive if and only if $\cos^2\theta\ge 1/5$. In this parameter range,
\begin{align}
\rho_S(t_2)
    = \Lambda_{t_2:t_1}[\rho_S(t_1)]
\end{align}
with a CPTP propagator. Hence, the two-step dynamics are CP-divisible for $\cos^2\theta\ge1/5$:
\begin{align}
\Lambda_{t_2:t_0} = \Lambda_{t_2:t_1}\Lambda_{t_1:t_0}.
\end{align}

To determine whether the process is Markovian, consider the full two-step process tensor $T_{2:0}$, defined such that for any pair of Choi matrices $C_0$ and $C_1$ representing operations at $t_0$ and $t_1$,
\begin{align}
\rho_S(t_2)
    = \mathrm{Tr}_{\text{aux}}
        \bigl[T_{2:0}(C_1 \otimes C_0)\bigr].
\end{align}
If the process were Markovian, $T_{2:0}$ would factorize as
\begin{align}
T_{2:0}^{\mathrm{Markov}}
    = C_{t_2:t_1} \otimes C_{t_1:t_0}.
\label{eq:MarkovFactorization}
\end{align}
Direct evaluation of $T_{2:0}$ for the correlated initial state~\eqref{eq:CorrEnv} shows that cross-terms appear involving both $C_0$ and $C_1$ simultaneously, reflecting the propagation of initial $E_1$ and $E_2$ correlations through the dynamics. As a result,
\begin{align}
T_{2:0}\;\neq\;C_{t_2:t_1} \otimes C_{t_1:t_0}.
\end{align}
Therefore, the process has non-zero temporal correlations and is non-Markovian, even though it is CP-divisible at the level of two-time maps (for $\cos^2\theta\ge1/5$). This example explicitly demonstrates that CP-divisible dynamics do not imply a Markovian process tensor. Initial correlations in the environment lead to multi-time memory that is invisible to any analysis based solely on two-time CPTP maps. Note also that, since $\rho_{E_1E_2}$ is a convex mixture of the product states $\ket{kk}$, the process is a mixture of two Markovian processes labeled by $k$: its memory is classical in the sense of Sec.~\ref{sec:QC_QNM} (cf.\ Appendix~\ref{exam:mixing}).
\begin{figure}
\centering
\includegraphics[width=0.85\linewidth]{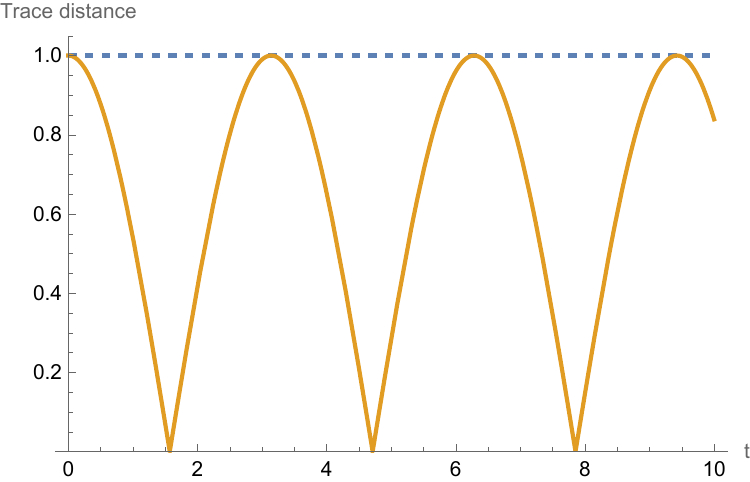}
\caption{Trace distance between the evolved states $\rho_\pm=\ket{\pm}\bra{\pm}$ for a qubit with Hamiltonians $H_i=\tfrac{\omega_i}{2}\sigma_z$ ($\omega_1=1$, $\omega_2=3$).
The dashed line corresponds to each individual Markovian unitary evolution, for which the trace distance is constant, while the solid curve shows the classically mixed dynamics $\Phi_t=\tfrac12(\Phi_t^{(1)}+\Phi_t^{(2)})$, exhibiting non-monotonic behavior.}
\label{fig:placeholder}
\end{figure}

\subsection{Example: classical mixing induces non-genuine non-Markovianity}
\label{exam:mixing}
We explicitly analyze the evolution of the state under a single Markovian unitary channel and under a classically mixed unitary dynamics, and compute the corresponding trace distance in both cases. Consider the initial state
\begin{align}
\ket{+}=\frac{1}{\sqrt{2}}(\ket{0}+\ket{1}),
\qquad
\rho_+(0)=\frac12
\begin{pmatrix}
1 & 1\\
1 & 1
\end{pmatrix}.
\end{align}
First, consider the Hamiltonian
\begin{align}
H_1=\frac{\omega_1}{2}\sigma_z,
\end{align}
which generates the unitary evolution
\begin{align}
U_1(t)=e^{-iH_1 t}
=
\begin{pmatrix}
e^{-i\omega_1 t/2} & 0\\
0 & e^{i\omega_1 t/2}
\end{pmatrix}.
\end{align}
The evolved state is
\begin{align}
\rho_+^{(1)}(t)
=
U_1(t)\rho_+(0)U_1^\dagger(t)
=
\frac12
\begin{pmatrix}
1 & e^{-i\omega_1 t}\\
e^{i\omega_1 t} & 1
\end{pmatrix}.
\end{align}
Thus, the diagonal terms remain invariant, while the coherences acquire opposite phases,
\begin{align}
\rho_{01}^{(1)}(t)&=\tfrac12 e^{-i\omega_1 t}, &
\rho_{10}^{(1)}(t)&=\tfrac12 e^{i\omega_1 t}.
\end{align}
If we consider the orthogonal state $\rho_-=\ket{-}\!\bra{-}$, its evolution under $U_1(t)$ yields
\begin{align}
\rho_-^{(1)}(t)
=
\frac12
\begin{pmatrix}
1 & -e^{-i\omega_1 t}\\
- e^{i\omega_1 t} & 1
\end{pmatrix}.
\end{align}
The difference between the two evolved states is
\begin{align}
\rho_+^{(1)}(t)-\rho_-^{(1)}(t)
=
\begin{pmatrix}
0 & e^{-i\omega_1 t}\\
e^{i\omega_1 t} & 0
\end{pmatrix},
\end{align}
which has eigenvalues $\pm 1$. The trace distance, therefore, reads
\begin{align}
D^{(1)}(t)
=
\frac12
\bigl\|
\rho_+^{(1)}(t)-\rho_-^{(1)}(t)
\bigr\|_1
=
1,
\end{align}
and is constant in time, confirming the absence of information backflow for the individual unitary evolution.

We now consider the classically mixed map
\begin{align}
\Phi_t=\frac12\,\Phi_t^{(1)}+\frac12\,\Phi_t^{(2)},
\end{align}
where $\Phi_t^{(i)}(\rho)=U_i(t)\rho U_i^\dagger(t)$ and
$H_i=\tfrac{\omega_i}{2}\sigma_z$ with $\omega_1\neq\omega_2$.
Applying $\Phi_t$ on $\rho_+(0)$, again, does not change the diagonal terms, while the off-diagonal terms change to
\begin{align}
\rho_{01}(t)
&=
\frac14\!\left(e^{-i\omega_1 t}+e^{-i\omega_2 t}\right),
&
\rho_{10}(t)
&=
\rho_{01}(t)^{*}.
\end{align}
Defining $\Delta\omega=\omega_1-\omega_2$, their magnitude becomes
\begin{align}
|\rho_{01}(t)|=|\rho_{10}(t)|
=
\frac12
\left|
\cos\!\left(\frac{\Delta\omega\, t}{2}\right)
\right|.
\end{align}

The corresponding evolved state $\rho_-(t)$ is obtained analogously, with opposite signs in the coherences. The difference
$\rho_+(t)-\rho_-(t)$ has eigenvalues
\begin{align}
\pm
\left|
\cos\!\left(\frac{\Delta\omega\, t}{2}\right)
\right|,
\end{align}
yielding the trace distance
\begin{align}
D(t)
=
\frac12
\bigl\|
\rho_+(t)-\rho_-(t)
\bigr\|_1
=
\left|
\cos\!\left(\frac{\Delta\omega\, t}{2}\right)
\right|.
\end{align}
This quantity is non-monotonic in time and exhibits revivals (see Fig.~\ref{fig:placeholder}), indicating information backflow induced solely by the classical mixing of Markovian unitary evolutions. There is no quantum contribution to the backflow of information.

\subsection{Examples: temporal correlations and non-Markovianity}\label{exm:tem}

\begin{example}[Identity channel on a qubit]\label{ex:sot_id}
For $\mathcal{E}=\mathrm{id}$, $J[\mathrm{id}]=\sum_{ij}\ket{i}\!\bra{j}\otimes\ket{j}\!\bra{i}=\mathrm{SWAP}$. With $\rho=\mathds{1}/2$,
\begin{align}
    \psi=\tfrac12\,\mathrm{SWAP}=\tfrac14\sum_{\mu=0}^{3}\sigma_\mu\otimes\sigma_\mu .
\end{align}
SWAP has eigenvalue $+1$ on the triplet and $-1$ on the singlet, so $\psi$ has spectrum $\{\tfrac12,\tfrac12,\tfrac12,-\tfrac12\}$.
The dynamics is unitary, hence Markovian in every sense of Sec.~\ref{Defs_Markov}, yet $\psi$ has a negative eigenvalue. This shows that negativity of $\psi$ is a witness of quantum temporal correlations of the system alone and is not, by itself, a witness of memory.
\end{example}

\begin{example}[Noise restores positivity]\label{ex:sot_dep}
Let $\mathcal{E}_p(\sigma)=(1-p)\sigma+p\Tr(\sigma)\mathds{1}/2$ and $\rho=\mathds{1}/2$. Then $J[\mathcal{E}_p]=(1-p)\,\mathrm{SWAP}+\tfrac p2\,\mathds{1}\otimes\mathds{1}$ and $\psi=\tfrac12J[\mathcal{E}_p]$, with eigenvalues
\begin{align}
    \tfrac12-\tfrac p4\ \ (\text{triplet}),\qquad -\tfrac12+\tfrac{3p}{4}\ \ (\text{singlet}).
\end{align}
Hence $\psi\ge 0$ if and only if $p\ge 2/3$. For $p=1$ one recovers the product state $\mathds{1}/2\otimes\mathds{1}/2$. This is consistent with the observation of Ref.~\cite{Fullwood2022} that sufficient noise on the channel and the input makes $\psi$ a genuine state, and with the connection of the Jordan product to the incompatibility of channels~\cite{Girard2021}.
\end{example}

\begin{example}[Singlet and the identity channel]\label{ex:sot_singlet}
For $\rho=\mathds{1}/2$, Eq.~\eqref{eq:sot_inverse} reduces to $X_\tau=2\tau$.

For $\tau=\tfrac12\mathrm{SWAP}$, this gives $X_\tau=\mathrm{SWAP}=J[\mathrm{id}]$, so $\mathcal{E}_\tau=\mathrm{id}$, recovering Example~\ref{ex:sot_id}. This $\tau$ is a state over time but not a state.

By contrast, the singlet $\tau=\proj{\psi^-}=\tfrac12(\mathds{1}-\mathrm{SWAP})$ is a genuine state with maximally mixed marginals. It gives $X_\tau=\mathds{1}-\mathrm{SWAP}$ and $\mathcal{E}_\tau(\sigma)=\Tr(\sigma)\mathds{1}-\sigma=\sigma_y\sigma^{T}\sigma_y$, i.e., the Bloch inversion $\vec r\mapsto-\vec r$. Being a transposition followed by a unitary, this map is positive and trace preserving but not completely positive. Indeed, $X_\tau^{T_{S_0}}=\mathds{1}-\sum_{i,j}\ket{ii}\!\bra{jj}$ has eigenvalue $-1$ on $(\ket{00}+\ket{11})/\sqrt2$. Hence, the singlet is a bipartite state, but not the time-evolved state of any quantum channel.

This asymmetry between spatial and temporal correlations has no classical counterpart~\cite{Fullwood2025}. Classically, every joint distribution $P(x,y)$ with $p(x)>0$ defines the stochastic map $P(y|x)=P(x,y)/p(x)$.
\end{example}

\end{document}